\documentclass{siamltex}

\usepackage{amsmath} 
\usepackage{amssymb} 

\input psfig.sty

\newcommand{\pl}[2]{\frac{\partial#1}{\partial#2}}

\newcommand{\p}{\partial} 
 
\newcommand{\Og}{\Omega} 
\newcommand{\fl}[2]{\frac{#1}{#2}} 
\newcommand{\dt}{\delta} 
\newcommand{\tm}{\times} 
 
\newcommand{\nn}{\nonumber} 
\newcommand{\ap}{\alpha} 
\newcommand{\bt}{\beta} 
 
\newcommand{\Gm}{\Gamma} 
\newcommand{\gm}{\gamma} 
 
\newcommand{\tp}{{\tilde{\phi}}}
\newcommand{\tht}{\theta} 
\newcommand{\ift}{\infty} 
\newcommand{\vep}{\varepsilon}

\newcommand{\btd}{\nabla} 
\newcommand{\btu}{\Delta}

\renewcommand{\theequation}{\arabic{section}.\arabic{equation}} 
\newcommand{\be}{\begin{equation}} 
\newcommand{\ee}{\end{equation}} 
\newcommand{\ba}{\begin{array}} 
\newcommand{\ea}{\end{array}} 
\newcommand{\bea}{\begin{eqnarray}} 
\newcommand{\eea}{\end{eqnarray}} 
\newcommand{\beas}{\begin{eqnarray*}} 
\newcommand{\eeas}{\end{eqnarray*}} 
\newcommand{\dpm}{\displaystyle} 
\newtheorem{remark}{Remark}[section]

\newcommand{\um}{\mu} 
 
\newcommand{\bx}{{\bf x} } 
\newcommand{\tu}{{\tilde u}}
\newcommand{\Eb}{{E_\beta}} 
\newcommand{\Ez}{{E_0}} 
\newcommand{\Et}{{\tilde{E}}} 
\newcommand{\Vt}{{\tilde{V}_n(\bx)}} 
\newcommand{\Etb}{{\tilde E_\beta}} 

\title{Computing the ground state solution of
Bose-Einstein condensates\\ by a normalized gradient flow} 

\author{ Weizhu Bao \thanks{Department of Computational Science,
National University of Singapore, Singapore 117543,
({\tt bao@cz3.nus.edu.sg}).
Research is supported  by the National University of 
Singapore grant No. R-151-000-027-112. }
\and 
 Qiang Du \thanks{Department of Mathematics,  Penn State University,
 University Park, PA 16802, USA ({\tt qdu@math.psu.edu}).
Research is supported in part by NSF DMS-0196522.}
}
 
\date{} 

\begin{document} 

\maketitle 
 
\begin{abstract} 
In this paper, we prove the energy diminishing of a normalized gradient flow
which provides a mathematical justification of the imaginary time method 
used in physical literatures to compute the ground state solution 
of Bose-Einstein condensates (BEC).  We also investigate
the energy diminishing  property  
for the discretization of the 
  normalized gradient flow. Two numerical methods are
proposed for such discretizations: one is the
 backward Euler centered finite difference (BEFD), 
the other one is an explicit
time-splitting sine-spectral (TSSP) method. 
Energy diminishing  for BEFD and TSSP
for linear case, and monotonicity for BEFD for both linear and
nonlinear cases are proven.
Comparison between the two methods and existing
methods, e.g. Crank-Nicolson finite difference (CNFD) 
or forward Euler finite difference (FEFD), 
shows that BEFD and TSSP are 
much better in terms of preserving
energy diminishing property of the normalized gradient flow.
Numerical results in 1d, 2d and 3d with magnetic trap confinement potential,
 as well as a potential of a stirrer
corresponding to a far-blue detuned Gaussian laser beam are reported
to demonstrate the effectiveness of  BEFD and TSSP methods. 
Furthermore we observe that the normalized gradient flow can also be applied 
directly to compute the first excited state 
solution in BEC when the initial data is 
chosen as an odd function.
\end{abstract}


 
\pagestyle{myheadings}
\markboth{W. Bao and Q. Du }{Computing the ground state of BEC}

\section{Introduction}\label{si} 
\setcounter{equation}{0} 
 
Since the first experimental realization of Bose-Einstein 
condensates (BEC) in dilute weakly interacting gases the 
nonlinear Schr\"{o}dinger equation (NLS), also called 
Gross-Pitaevskii equation (GPE) \cite{LL,Pit},
 has been used extensively 
to describe the single particle properties of BECs. The results 
obtained by solving the NLS showed excellent agreement with most 
of the experiments (for a review see 
\cite{Anglin,Dalfovo,Cornell}). In fact, up to now there have been 
very few experiments in ultracold dilute bosonic gases which could 
not be described properly by using theoretical methods based on 
the NLS \cite{NGPEExp,NGPETheo}. 
 
   There has been a series of recent studies which deal
with the numerical solution of the time-independent GPE for ground state
and the time-dependent GPE for finding the dynamics of a BEC. 
For numerical solutions of time-dependent GPE, Bao et al. 
\cite{Bao1,BJP,Bao3,Bao4}
presented a time-splitting spectral method,  Ruprecht et al.  \cite{Rup}
used the Crank-Nicolson finite difference method to compute the ground 
state solution and dynamics of GPE, Cerimele et. al. 
\cite{Cerim} proposed  a particle-inspired scheme. 
For ground state solution 
of GPE, Edwards et al. presented a Runge-Kutta type method 
and used it to solve 1d and 3d with spherical symmetry time-independent GPE
\cite{Edwards}. Adhikari \cite{Adh,Adh1} 
used this approach to get the ground state solution
of GPE in 2d with radial symmetry. Bao et al. \cite{Bao} proposed a
method by directly minimizing the energy  functional. 
Other approaches include 
an explicit imaginary-time 
algorithm used  by Chiofalo et al. \cite{Tosi}, a direct
inversion in the iterated subspace (DIIS) used by Schneider et al. 
\cite{Feder}, and a simple analytical 
type method proposed by Dodd \cite{Dodd}. In fact, 
one of the fundamental problems in numerical simulation of BEC
is to compute the ground state solution.

 We consider the NLS equation \cite{Bao3,Sulem}
\bea 
\label{sdge} 
&&i\; \psi_t=-\fl{1}{2}\;\btu \psi+ V(\bx)\; \psi +\bt 
|\psi|^{2}\psi, \qquad t>0, \qquad \bx\in \Og\subseteq{\mathbb R}^d, \\
\label{sdge1} 
&&\psi(\bx,t)= 0, \qquad \bx \in \Gm=\p\Og, \quad t\ge0;
\eea 
where $\Og$ is a subset of ${\mathbb R}^d$
and $V({\bf x})$ is a real-valued potential whose shape is determined 
by the type of system under investigation, and $\bt$ positive/negative 
corresponds to the defocusing/focusing NLS. 
(\ref{sdge}) is known in BEC as the Gross-Pitaevskii 
equation (GPE) \cite{Pit} where $\psi$ is the macroscopic wave function 
of the condensate, $t$ is time, ${\bf x}$ is the spatial 
coordinate and $V({\bf x})$ is a trapping potential which usually 
is harmonic and can thus be written as $V({\bf x})=\fl{1}{2} 
\left(\gm_1^2 x_1^2 +\cdots+\gm_d^2 x_d^2\right)$ with $\gm_1, 
\cdots, \gm_d> 0$. Two important invariants of (\ref{sdge}) are 
the {\bf normalization of the wave function} 
\begin{equation} \label{mass} 
N(\psi)=\int_{\Og}\; |\psi({\bf x}, t)|^2\; d{\bf x}=1, 
\qquad t\ge 0 
\end{equation} 
and the {\bf energy} 
\begin{equation} \label{energy} 
\Eb(\psi) = \int_{\Og}\; \left[\fl{1}{2}|\btd \psi({\bf x}, 
t)|^2 
+V(\bx)|\psi(\bx,t)|^2+\fl{\bt}{2}|\psi(\bx,t)|^{4}\right]\; 
d\bx, \quad t\ge 0. 
\end{equation} 
 
To find a stationary solution of (\ref{sdge}), we write 
\be
\label{stat}
\psi(\bx,t)=e^{-i\mu t} \phi(\bx),
\ee
where $\mu$ is the chemical potential
of the condensate and $\phi$ a real function independent of time.
Inserting into (\ref{sdge})
gives the following equation for $\phi(\bx)$
\bea
\label{gss}
&&\mu\; \phi(\bx)=-\fl{1}{2}\btu \phi(\bx)+
V(\bx)\;\phi(\bx)
+ \bt |\phi(\bx)|^2\phi(\bx), \qquad \bx\in \Og, \\
\label{gss1}
&&\phi(\bx)= 0, \qquad \bx \in \Gm;
\eea
under the normalization condition 
\be
\label{normgg}
\int_{\Og} \; |\phi(\bx)|^2\;d\bx=1.
\ee
This is a nonlinear eigenvalue problem under a constraint and any eigenvalue
$\mu$ can be computed from its corresponding eigenfunction $\phi$ by
\begin{eqnarray}
\mu=\mu_\bt(\phi) &= &\int_\Og\left[\fl{1}{2}\left|\btd \phi(\bx)\right|^2
+V(\bx)\left|\phi(\bx)\right|^2 +\bt \left|\phi(\bx)
\right|^4\right]d\bx\nonumber\\
& = &E_\bt(\phi)+\int_\Og \; \fl{\bt}{2}\left|\phi(\bx)\right|^4\;d\bx.
\label{engvf}
\end{eqnarray}
The non-rotating Bose-Einstein condensate
ground state solution $\phi_g(\bx)$ is a real nonnegative function
found by minimizing the energy $\Eb(\phi)$ under the constraint 
(\ref{normgg}) \cite{Lieb}. 
In physical literatures \cite{Adu,Tosi2,Tosi}, 
this minimizer 
was obtained by applying an imaginary time (i.e. $t\to -it$) 
in (\ref{sdge}) 
and evolving a normalized gradient flow (see details in 
the next section). In fact, it is easy to show that the minimizer of
$\Eb(\phi)$ under the constraint (\ref{normgg}) is an eigenfunction of 
(\ref{gss}). 

The aim of this paper is to prove energy diminishing of the 
normalized gradient flow and present two new 
numerical methods to discretize the normalized gradient flow.
This gives a mathematical justification of 
the imaginary time method which is widely used in 
physical literatures to compute the ground state solution of BEC.
Energy diminishing of the discretization of the normalized 
gradient flow is also proven. Extensive numerical results
are reported to demonstrate the effectiveness of
our new methods.
 
The paper is organized as follows. In section \ref{sNGF} we 
prove energy diminishing of the normalized gradient flow
and its discretized version. 
 In section \ref{smethod} we propose two numerical discretizations
for the normalized gradient flow. In section \ref{sne} 
numerical comparison between the two methods and existing methods,
as well as applications of the two methods for 1d, 2d and 3d ground
state solution of BEC, are reported.
Finally in section \ref{sc} some conclusions are drawn. 
Throughout we adopt the standard notation for Sobolev spaces.

Before we end the introduction, let us note that the NLS is also used in 
nonlinear optics, e.g., to describe the propagation of an intense 
laser beam through a medium with a Kerr nonlinearity 
\cite{FibichP, Sulem} where $\psi=\psi({\bf x},t)$ 
describes the electrical field amplitude, $t$ is the spatial 
coordinate in the direction of propagation, ${\bf x} =(x_1,\cdots, 
x_d)^T$ is the transverse spatial coordinate and $V({\bf x})$ is 
determined by the index of refraction.

\section{Normalized gradient flow} 
\label{sNGF} 
\setcounter{equation}{0} 
 
In this section we prove energy diminishing of a 
normalized gradient flow and its discretized version.

\subsection{Energy diminishing}

  Consider the gradient flow
\bea
\label{gf1}
&&u_t = \fl{1}{2}\btu u - V(\bx) u -\bt\; |u|^2 u, \qquad t>0, \quad 
\bx\in \Og,\\
\label{gf2}
&&u(\bx,0)=u_0(\bx), \qquad \bx \in \Og,\\
\label{gf3}
&&u(\bx,t)= 0, \qquad \bx \in \Gm, \quad t\ge0;
\eea
where  $\|u_0\|=1$. Here we adopt the norm by 
$\|\cdot\|=\|\cdot\|_{L^2(\Og)}$ and denote
$\|\cdot\|_{L^m}=\|\cdot\|_{L^m(\Og)}$ with $m$  an integer. 
Let 
\be
\label{nu}
\tu(\cdot, t)=\fl{u(\cdot,t)}{\|u(\cdot,t)\|}, \qquad t\ge0. 
\ee
Then, it is easy to establish the following basic facts:

\begin{theorem}\label{energyd}
Suppose $V(\bx)\ge0$ for all $\bx \in \Og$, $\bt\ge 0$ and
$\|u_0\|=1$, then

(i). $\|u(\cdot, t)\| \le \|u(\cdot, 0)\| =\|u_0\| =1$ for $0\le t<\ift$.

(ii). For any $\bt\geq 0$,
\be
\label{energydg}
\Eb(u(\cdot,t))\le  \Eb(u(\cdot,t')), \qquad 0\le t'<t<\ift.
\ee

(iii). For $\bt=0$, 
\be
\label{energydl}
\Ez(\tu(\cdot,t))\le \Ez(\tu(\cdot,0)) = \Ez(u_0), \qquad 0\le t<\ift.
\ee
\end{theorem}

\noindent {\bf Proof:} \ (i). From (\ref{gf1}) and (\ref{gf3}), integration
by parts, we get
\bea
\label{du2}
\fl{d}{dt}\|u\|^2&=&\fl{d}{dt}\int_\Og u^2\; d\bx =\int_\Og 2 u\; u_t\; d\bx
=\int_\Og 2 u\left[ \fl{1}{2}\btu u- V(\bx) u -\bt |u|^2u\right] \; d\bx \nn\\
&=&-2\int_\Og\left[\fl{1}{2}|\btd u|^2 +V(\bx) u^2 +\bt u^4\right]\;d\bx
\le 0, \qquad 0\le t<\ift.
\eea 
This implies the result in (i).

 (ii). From (\ref{gf1}), (\ref{gf3}) and 
(\ref{energy}) with $\psi=u$, integration
by parts, we get
\bea
\label{du4}
\fl{d}{dt} \Eb(u)&=& 2 \int_\Og 
\left[\fl{1}{2}\btd u \btd u_t+ 
u_{t}( V(\bx) u +\bt |u|^2 u)\right] \; d\bx \nn\\
&=& - 2 \int_\Og 
u_{t}\left[ \fl{1}{2}\btu u - V(\bx) u -\bt |u|^2 u\right] \; d\bx \nn\\
&=&- 2 \int_\Og |u_t|^2 \;d\bx \le 0, \qquad 0\le t<\ift.
\eea
This implies the result in (ii).

(iii). 
From (\ref{energy}) with $\psi=\tu$ and $\bt=0$, 
(\ref{gf1}), (\ref{gf3}), (\ref{nu}), 
(\ref{du2}) and (\ref{du4}), 
integration by parts and  Schwartz inequality, we obtain
\bea
\label{dE}
\lefteqn{\fl{d}{dt}\Ez(\tu)=\fl{d}{dt}\int_\Og\left[\fl{|\btd u|^2}{2\|u\|^2}+
\fl{V(\bx)u^2}{\|u\|^2} \right]\;d\bx}\nn\\ [2mm]
&=&2\int_\Og\left[\fl{\btd u\cdot \btd u_t}{2\|u\|^2}+
\fl{V(\bx)u\;u_t}{\|u\|^2} \right]d\bx-\left(\fl{d}{dt}\|u\|^2\right)\;
\int_\Og\left[\fl{|\btd u|^2}{2\|u\|^4}+
\fl{V(\bx)u^2}{\|u\|^4} \right]d\bx \nn\\
&=&2\int_\Og \fl{\left[-\fl{1}{2}\btu u +V(\bx)u \right]u_t}
{\|u\|^2}\;d\bx-\left(\fl{d}{dt}\|u\|^2\right)\int_\Og
\fl{\fl{1}{2}|\btd u|^2 +V(\bx) u^2 }{\|u\|^4}\;d\bx\nn\\
&=&-2\fl{\|u_t\|^2}{\|u\|^2}+\fl{1}{2\|u\|^4}
\left(\fl{d}{dt}\|u\|^2\right)^2 \nn\\
&=&\fl{2}{\|u\|^4}\left[\left(\int_\Og u\; u_t \; d\bx\right)^2-
\|u\|^2 \|u_t\|^2\right]
 \nn\\
&\le& 0\; , \qquad 0\le t<\ift.
\eea 
This implies (\ref{energydl}). \hfill $\Box$

\vskip20pt

\begin{remark} \label{edm}
The property 
(\ref{energydg}) is often referred as the energy diminishing
property of the gradient flow. It is interesting to note
that (\ref{energydl}) implies that
the energy diminishing property is preserved even in
the normalized gradient flow when $\bt=0$, that is, for
linear evolution equations.
\end{remark}

\vskip20pt

\begin{remark} \label{finitet0}
When $\bt>0$, the solution of
(\ref{gf1})-(\ref{gf2})  may not preserve the 
normalized energy diminishing property
$$\Eb(\tu(\cdot,t))\le  \Eb(\tu(\cdot,t')), \qquad 0\le t'<t<\ift.$$
In fact,
we solve 
(\ref{gf1})-(\ref{gf2}) 
in 1d with $\Og={\mathbb R}$ and
$V(x)=x^2/2$ numerically by the time-splitting spectral method (
see details in the next section) for the initial condition
$u_0(x) =\dpm (\pi/2)^{-1/4}\; e^{-x^2}$.
Figure \ref{fig1t} shows, for different $\bt$, the
energy $\Eb(\tu(\cdot, t))=\Eb\left(u(\cdot,t)/\|u(\cdot,t)\|\right)$
under mesh size $h=1/32$
and time step $k=0.0001$. From the figure, we can see that $\Eb(\tu)$ 
diminishing for $0\le t<\ift$ when $\bt=0$. But  when $\bt>0$,
we have $\Eb(\tu)$ diminishing only for 
$0\le t\le t_0$ with some finite $t_0<\ift$. 
\end{remark} 

\begin{figure}[hb] \label{fig1t}
\centerline{\psfig{figure=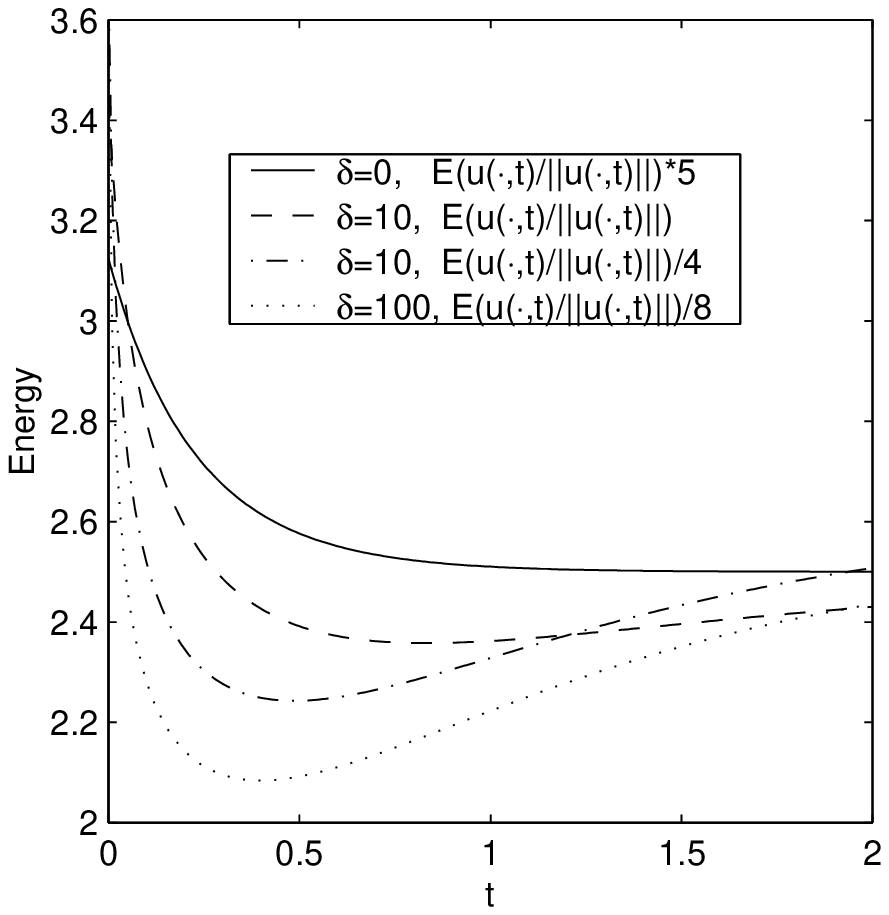,height=7.5cm,width=7.5cm,angle=0} }  
\caption{$\Eb(\tu)$ as a function of time in Remark \ref{finitet0} 
for  different $\bt$ (labeled as $\dt$).}
\end{figure} 

\subsection{Normalized gradient flow}

Consider the following continuous 
normalized gradient flow
\bea
\label{nkngf1}
&&\phi_t = \fl{1}{2}\btu \phi - V(\bx) \phi -\bt\; |\phi|^2\phi
+\um_\phi(t)\phi, \qquad 
\bx\in \Og, \quad t\ge 0,\\
\label{nkngf2}
&&\phi(\bx,t)= 0, \qquad \bx \in \Gm,\\
\label{nkngf3}
&&\phi(\bx,0)=\phi_0(\bx), \qquad \bx \in \Og.
\eea
In fact, the right hand side of  (\ref{nkngf1}) is the same
as (\ref{gss}) if we view $\mu_\phi(t)$ 
as a Lagrange multiplier for the 
constraint (\ref{normgg}). 
It readily follows that
\be
\label{sgt}
\um_\phi(t)=\fl{1}{\|\phi(\cdot,t)\|^2}\int_{\Og}
\left[\fl{1}{2}|\btd \phi(\bx,t)|^2+V(\bx)|\phi|^2(\bx,t)+
\bt|\phi|^4(\bx,t)\right]d\bx\; .
\ee
Furthermore for the above normalized gradient flow, 
as observed in \cite{Adu,Du2},
the solution of (\ref{nkngf1}) also satisfies the following theorem:



 \begin{theorem}\label{edhh}
Suppose $V(\bx)\ge0$ for all $\bx \in\Og$, $\bt\ge0$ and 
$\|\phi_0\|=1$. Then the normalized gradient flow 
(\ref{nkngf1})-(\ref{nkngf3}) is normalization conservation and
energy diminishing, i.e.
\bea
\label{ncphi}
&&\|\phi(\cdot,t)\|^2=\int_\Og \phi^2(\bx,t)\; d\bx =
\|\phi_0\|^2=1, \qquad t\ge0,\\
\label{edcngf}
&&
\fl{d}{dt}\Eb(\phi)=- 2\left\|\phi_t(\cdot,t)\right\|^2\le 0\;, 
\qquad t\ge0,   
\eea
which in turn implies
$$
\Eb(\phi(\cdot, t_1))\ge \Eb(\phi(\cdot,t_2)), \qquad 0\le t_1\le t_2<\ift.
$$
\end{theorem}

\noindent {\bf Proof:} \ Multiplying both sides of (\ref{nkngf1}) by 
$\phi$, integrating over $\Og$, integration by parts and notice 
(\ref{sgt}), (\ref{nkngf2}), we obtain
\bea
\label{ncng}
\lefteqn{\fl{1}{2}\fl{d}{dt}\int_\Og |\phi(\bx,t)|^2\;d\bx = 
\int_\Og \phi\; \phi_t\; d\bx  }\nn\\[2mm]
& = &  
\int_\Og\left[\fl{1}{2}\btu \phi - V(\bx) \phi -\bt\; \phi^3
+\um_\phi(t)\phi\right]\phi\;d\bx 
 \nn\\[2mm]
&=&-\int_\Og \left[\fl{1}{2}|\btd \phi(\bx,t)|^2+V(\bx)\phi^2(\bx,t)+
\bt\phi^4(\bx,t)\right]d\bx +\um_\phi(t) \|\phi(\cdot,t)\|^2 \nn\\
&=&0, \qquad t\ge0.
\eea
This implies the normalization conservation (\ref{ncphi}).

Next, direct calculation shows
\bea
\label{decngf}
\fl{d}{dt}\Eb(\phi)&=&
\int_\Og \left[
\fl{1}{2}\btd\phi \cdot \btd \phi_t +V(\bx) \phi \phi_t +\bt \phi^3 \phi_t
\right]d\bx \nn \\ 
&=&2 
\int_\Og \left[-\fl{1}{2}\btu \phi +V(\bx) \phi +\bt \phi^3\right]\phi_t
\; d\bx \nn 
  \\
&=&2 
\int_\Og \left[-\phi_t(\bx,t) +\mu_\phi(t) \phi(\bx,t)\right] \phi_t\;
d\bx \nn 
  \\ 
&=&- 2\|\phi_t(\cdot, t) \|^2 
+\mu_\phi(t)\; \fl{d}{dt} \int_\Og |\phi(\bx,t)|^2\;d\bx
\nn \\
&=&- 2\|\phi_t(\cdot, t) \|^2 \;, \qquad t\ge0,
\eea
since $\mu_\phi(t)$ is always real and
$$
\fl{d}{dt} \int_\Og  |\phi(\bx,t)|^2\;d\bx =0
$$
due to the normalization conservation. Thus, we easily get
$$
\Eb(\phi(\cdot, t_1))\ge \Eb(\phi(\cdot,t_2)), \qquad 0\le t_1\le t_2<\ift
$$
for the solution of (\ref{nkngf1}).
\hfill $\Box$

\vskip20pt
\begin{remark}
We see from the above theorem that
the energy diminishing property is preserved in the continuous
dynamic system (\ref{nkngf1}). 
\end{remark} 

Using argument similar to that in \cite{LD,Simon}, we may also
get as $t\to\ift$, $\phi$ approaches to a steady state solution
which is a critical point of the energy.  In non-rotating BEC, 
it has a unique real valued nonnegative 
ground state solution $\phi_g(\bx)\ge0$ for all $\bx\in\Og$  \cite{Lieb}.
We choose the 
initial data $\phi_0(\bx)\ge0$ for $\bx\in\Og$, e.g. 
the ground state solution of linear Schr\"{o}dinger equation
with a harmonic oscillator potential \cite{Bao,Bao3}.
Under this kind of initial data, 
the ground state solution $\phi_g$ and its corresponding chemical
potential $\mu_g$ can be obtained from 
the steady state solution of the normalized gradient flow 
(\ref{nkngf1})-(\ref{nkngf3}), i.e. 
\be
\label{gstlim}
\phi_g(\bx)=\lim_{t\to\ift} \phi(\bx,t),\quad \bx\in\Og,
\qquad  
\mu_g =\mu_\bt(\phi_g)=\Eb(\phi_g)+\fl{\bt}{2}\int_\Og \phi_g^4(\bx)\; d\bx.
\ee

\subsection{Normalized gradient flow via splitting}

Various algorithms for computing the steady state solutions
of the normalized gradient flows have been studied in the literature.
For instance, 
second order in time discretization scheme that preserves
the norm normalization and energy diminishing properties were 
presented in \cite{Adu,Du2}. Perhaps one of the more popular
technique for dealing with the normalization constraint is
through the following construction:
choose a time sequence $0=t_0<t_1<t_2<\cdots<t_n<\cdots$ with
$\btu t_{n}=t_{n+1}-t_{n}>0$ and $k=\max_{n\ge 0} \; \btu t_n$. 
To adapt an algorithm for the solution of the
usual gradient flow to the case of normalized gradient flow,
it is natural to consider the following splitting (or projection)
scheme which was  widely used in physical literatures 
\cite{Du2,Tosi,Tosi2} for computing the ground state solution of BEC:
\bea
\label{ngf1}
&&\phi_t = \fl{1}{2}\btu \phi - V(\bx) \phi -\bt\; |\phi|^2\phi, \qquad 
\bx\in\Og,\quad t_n<t<t_{n+1}, \quad n\ge0,\\
\label{ngf4}
&&\phi(\bx,t)= 0, \qquad \bx \in \Gm,\\
\label{ngf2}
&&\phi(x,t_{n+1})\stackrel{\triangle}{=}
\phi(\bx,t_{n+1}^+)=\fl{\phi(\bx,t_{n+1}^-)}{\|\phi(\cdot,t_{n+1}^-)\|}, 
\qquad \bx\in \Og, \quad n\ge 0,\\
\label{ngf3}
&&\phi(\bx,0)=\phi_0(\bx), \qquad \bx \in \Og;
\eea
where $\phi(\bx, t_n^\pm)=\lim_{t\to t_n^\pm} \phi(\bx,t)$ and 
$\|\phi_0\|=1$. In fact, (\ref{ngf1}) is the same
as the original gradient flow (\ref{gf1}) which can thus be solved via
traditional techniques. The normalization of the
gradient flow is simply achieved by a normalization at
each discrete time step.

From Theorem \ref{energyd}, we get immediately

\begin{theorem}\label{edh}
Suppose $V(\bx)\ge0$ for all $\bx \in \Og$ and 
$\|\phi_0\|=1$.  For $\bt=0$,  the normalized
gradient flow is energy diminishing under any time step $k$ and
initial data $\phi_0$, i.e.
\be
\label{dphi}
\Ez(\phi(\cdot,t_{n+1}))\le \Ez(\phi(\cdot, t_n))\le \cdots
\le \Ez(\phi(\cdot,0))=\Ez(\phi_0), \;\; n=0,1,2,\cdots.
\ee


\end{theorem}

In fact, the normalized step (\ref{ngf2}) is equivalent to solve
the following ODE {\sl exactly}
\bea
\label{Ode1}
&&\phi_t(\bx,t) = \um_\phi(t,k) \phi(\bx,t), \qquad \bx\in\Og,
\quad t_n < t<t_{n+1}, \quad n\ge0,\\
\label{Ode2}
&&\phi(\bx,t_n^+)= \phi(\bx,t_{n+1}^-), \qquad   \bx\in\Og;
\eea
where 
\be
\label{sgtk}
\um_\phi(t,k)\equiv
\um_\phi(t_{n+1},\btu t_n) = -\fl{1}{2\; \btu t_n}
\ln \|\phi(\cdot,t_{n+1}^-)\|^2, 
\qquad t_n\le t\le t_{n+1}.
\ee
Thus the normalized gradient flow can be viewed as a first-order
splitting method for gradient flow with discontinuous coefficients: 
\bea
\label{nngf1}
&&\phi_t = \fl{1}{2}\btu \phi - V(\bx) \phi -\bt\; |\phi|^2\phi
+\um_\phi(t,k)\phi, \qquad 
\bx\in \Og, \quad t\ge 0,\\
\label{nngf2}
&&\phi(\bx,t)= 0, \qquad \bx \in \Gm,\\
\label{nngf3}
&&\phi(\bx,0)=\phi_0(\bx), \qquad \bx \in \Og.
\eea
Let $k\to 0$, 
we see that 
\be
\lim_{k\to0^+}\um_\phi(t,k)= \um_\phi(t)
=\fl{1}{\|\phi(\cdot,t)\|^2}\int_{\Og}
\left[\fl{1}{2}|\btd \phi(\bx,t)|^2+V(\bx)\phi^2(\bx,t)+
\bt\phi^4(\bx,t)\right]d\bx, 
\ee
which implies that  the problem of (\ref{nngf1})-(\ref{nngf3}) collapses to
(\ref{nkngf1}) as $k\to 0$.

\begin{remark}
As we noted earlier, the energy diminishing property in general
does not hold uniformly for all $\phi_0$ and all step size $k$.
Thus, we propose to consider a modified splitting step which simplifies
the computation and yet guarantees the monotonicity when it is 
discretized by BEFD further.
\end{remark}




\subsection{Semi-implicit time discretization}

To further discretize the equation,
we here consider the following semi-implicit time discretization
 scheme:

\bea
\label{ngfsp1}
&&\frac{\tp^{n+1}-\phi^n}{k} = 
\fl{1}{2}\btu \tp^{n+1} - V(\bx) \tp^{n+1} -\bt\; |\phi^n|^2\tp^{n+1}\; , 
\qquad 
\bx\in \Og, \\
\label{ngfsp2}
&&\tp^{n+1}(\bx)= 0, \qquad \bx \in \Gm,\\
&&\label{ngfsp3}
\phi^{n+1}(\bx)= \tp^{n+1}(\bx) /\|\tp^{n+1}\|\; ,
 \qquad \bx \in \Og\; .
\eea

Notice that since the equation (\ref{ngfsp1}) becomes linear, 
the solution at the new time step becomes relatively simple.

By defining
$$
\Vt=V(\bx)+ \bt |\phi^n(\bx)|^2\;, \qquad \bx \in \Og,
$$
we may rewrite (\ref{ngfsp1}) as
\be
\label{ngfspl1}
\frac{\tp^{n+1}-\phi^n}{k} = \fl{1}{2}\btu \tp^{n+1} - \Vt \tp^{n+1} \; .
\ee
In other words, in each discrete time interval, we may
view (\ref{ngfsp1}) as a discretization of 
a linear gradient flow with a modified potential $\Vt$.

We now first present the following lemma:

\begin{lemma}\label{ldh}
Suppose $\Vt \ge 0$ for all $\bx \in \Og$ and 
$\|\phi^n\|=1$.  Then,
\bea
\label{dphi1}
\int_\Og | \tp^{n+1}|^2 \;d\bx 
& \leq & 
\int_\Og \phi^n \;\tp^{n+1}\; d\bx, \\
\label{dphi2}
\int_\Og | \tp^{n+1}|^4\; d\bx 
& \leq & 
 \int_\Og |\phi^n|^2\; |\tp^{n+1}|^2\; d\bx. 
\eea
\end{lemma}

\noindent {\bf Proof:} \ Multiplying both sides of (\ref{ngfsp1}) by 
$\tp^{n+1}$, integrating over $\Og$, and applying integration by parts,
we obtain
$$
\int_\Og  \left( | \tp^{n+1}|^2 
-  \phi^n \tp^{n+1} \right) d\bx =  
- k \int_\Og \left[
\fl{1}{2}|\btd\tp^{n+1} |^2 +\Vt| \tp^{n+1}|^2 
\right]d\bx \leq 0 \; ,$$
which leads to (\ref{dphi1}).
Similarly, 
\bea
\int_\Og  | \tp^{n+1}|^2 
| \phi^n |^2  d\bx 
& =  &
\int_\Og  | \tp^{n+1}|^2 \left|
 \tp^{n+1} - \frac{k}{2} \btu \tp^{n+1}
+k \Vt \tp^{n+1} \right|^2  d\bx 
\nn\\[2mm]
& =  &
\int_\Og  | \tp^{n+1}|^2 \left[ | \tp^{n+1}|^2 -2 
\;\frac{k}{2}  \tp^{n+1}\btu \tp^{n+1} 
+ 2 k \Vt |\tp^{n+1}|^2   \right] d\bx \nn \\[2mm]
& & \qquad + \int_\Og  | \tp^{n+1}|^2 
\left| \frac{k}{2} \btu \tp^{n+1}  - k \Vt \tp^{n+1} \right|^2
 d\bx 
\nn\\[2mm]
& =  &
\int_\Og  | \tp^{n+1}|^2 \left[ | \tp^{n+1}|^2 
 + 3k |\btd \tp^{n+1}|^2 
+ 2 k \Vt |\tp^{n+1}|^2  \right] d\bx  \nn \\[2mm]
& & \qquad + \int_\Og  | \tp^{n+1}|^2 
 \left| \frac{k}{2} \btu \tp^{n+1}  - k \Vt \tp^{n+1} \right|^2
 d\bx 
\nn\\[2mm]
& \geq  &
\int_\Og  | \tp^{n+1}|^4 d\bx \;.
\eea
This implies (\ref{dphi2}).  \hfill $\Box$

\bigskip

Given a linear self-adjoint operator $A$ in a Hilbert space
$H$ with inner product $(\cdot,\cdot)$, 
and assume that $A$ is positive definite in the sense
that for some positive constant $c$,  
$(u,Au)\geq c (u, u)$ for any $u\in H$. 
We now present a simple lemma:

\begin{lemma}
For any $k>0$, and $(I+kA)u=v$, we have
\be
\label{gen1}
\frac{(u,Au)}{(u,u)}\leq
\frac{(v,Av)}{(v,v)} \; .
\ee
\end{lemma}

\noindent {\bf Proof:}\  Since $A$ is self-adjoint
and positive definite, by
H\"{o}lder inequality, we have for any $p,q\geq 1$ with
$p+q=pq$
$$
\left(u, A u\right)\leq 
\left(u,u\right)^{1/p}\left(u,A^{q}u\right)^{1/q}\; ,$$
which leads to 
$$
\left(u, A u\right)\leq 
\left(u,u\right)^{1/2}\left(u,A^{2}u\right)^{1/2}\; $$
and 
$$
\left(u, A u\right)\left(u,A^2u\right)
\leq 
\left(u,u\right)\left(u,A^3u\right) \;. $$
Direct calculation then gives
\bea
\lefteqn{\left(u,Au\right)\; \left((I+kA)u, 
(I+kA)u\right)}\nn\\[2mm]
&=&\left(u,Au\right)\; \left(u,u\right)
+2k \left(u, Au\right)^2+k^2 \left(u, Au\right)\left(u,A^2u\right)\nn\\
&\le&\left(u,Au\right)\; \left(u,u\right)+
2k \left(u, u\right)\; \left(u, A^2u\right)
+k^2\left(u, u\right)\; 
\left(u, A^3u\right) \nn\\
&=&\left(u,u\right)
\left((I+kA)u, A(I+kA)u\right).  
\eea
\hfill $\Box$

Let us defined a modified energy $\Et_{\phi^n}$ as
$$
\Et_{\phi^n}(u)=\int_\Og \left[\frac{1}{2} |\btd u|^2 
+ \Vt |u|^2\right] \; d\bx=\int_\Og \left[\frac{1}{2} |\btd u|^2 
+ V(\bx) |u|^2+\bt |\phi^n|^2|u|^2\right] \; d\bx \, ,
$$
we then get from the above lemma that

\begin{lemma}\label{ldhs}
Suppose $V(\bx) \ge 0$ for all $\bx \in \Og$, $\bt\ge 0$  and 
$\|\phi^n\|=1$.  Then,
\bea
\label{dsp1}
\Et_{\phi^n}(\tp^{n+1})\le \fl{\Et_{\phi^n}(\tp^{n+1})}{\|\tp^{n+1}\|}
=\Et_{\phi^n}\left(\fl{\tp^{n+1}}{\|\tp^{n+1}\|}\right)
=\Et_{\phi^n}(\phi^{n+1})
\leq \Et_{\phi^n}(\phi_n)\;. 
\eea
\end{lemma}

Using the inequality
(\ref{dphi2}), this in turn implies:

\begin{lemma}
Suppose $V(\bx) \ge 0$ for all $\bx \in \Og$ and $\bt\ge0$, then,
$$
\Etb(\tp^{n+1})\leq \Etb(\phi^n), 
$$
where 
\[\Etb(u)=\int_\Og \left[\frac{1}{2} |\btd u|^2 
+ V(\bx)|u|^2 +\bt |u|^4\right] \; d\bx \, .\]
\end{lemma}

\begin{remark}
For $\bt=0$, the energy diminishing property is preserved 
in the normalized gradient flow via splitting (\ref{ngf1})-(\ref{ngf3})
and semi-implicit time discretization (\ref{ngfsp1})-(\ref{ngfsp3}). 
For $\bt>0$, we could only justify the energy diminishing on 
a modified energy in two adjacent steps.
\end{remark}

\subsection{Discretized normalized gradient flow}

Consider a discretization for the normalized gradient glow 
(\ref{ngfsp1})-(\ref{ngfsp3}) 
(or a fully discretization of (\ref{nkngf1})-(\ref{nkngf3}))
\be
\label{dge}
\fl{\tilde {U}^{n+1}-U^n}{k} = -A \tilde {U}^{n+1}, 
\qquad U^{n+1}=\fl{\tilde {U}^{n+1}}{\|\tilde {U}^{n+1}\|},
\qquad n=0,1,2,\cdots;
\ee
where $U^n=(u_1^n, u_2^n, \cdots, u_{M-1}^n)^T$, $k>0$ is time step
and $A$ is an $(M-1)\tm (M-1)$ symmetric positive definite matrix.
We adopt the inner product,  norm  and energy of vectors
$U=(u_1,u_2,\cdots, u_{M-1})^T$ and $V=(v_1,v_2,\cdots, v_{M-1})^T$
as
\be
\label{normU}
(U,V)=U^T V=\sum_{j=1}^{M-1} u_j\; v_j, \quad \|U\|^2=U^T U=(U,U),\quad
\Ez(U)=U^T A U =(U, AU), 
\ee
respectively. 
Using the finite dimensional version of the
lemmas given in the previous subsection, we have

\begin{theorem} \label{deeg}
Suppose $\|U^0\|=1$ and $A$ is symmetric positive definite.
Then the discretized normalized
gradient flow (\ref{dge}) is energy diminishing, i.e.
\be
\label{edde}
\Ez\left(U^{n+1}\right)\le \Ez\left(U^n\right)\le \cdots
\le \Ez\left(U^0\right), \qquad n=0,1,2,\cdots.
\ee
Furthermore if $I+kA$ is an $M$-matrix \cite{Golub}, then $(I+kA)^{-1}$ 
is a nonnegative matrix (i.e. every entry in it 
is nonnegative). Thus the
flow is monotone, i.e. if $U^0$ is a non-negative vector,
then $U^n$ is also a non-negative vector for all $n\ge0$.
\end{theorem}


\begin{remark}\label{forw}
If a discretization for the normalized 
gradient flow (\ref{ngfsp1})-(\ref{ngfsp3})   reads
\be
\label{dgef}
\fl{\tilde{U}^{n+1}-U^n}{k} = -B U^{n}, 
\qquad U^{n+1}=\fl{\tilde {U}^{n+1}}{\|\tilde {U}^{n+1}\|}, 
\qquad n=0,1,2,\cdots.
\ee
Suppose $B$ is symmetric and 
positive definite and $\rho(kB)<1$ where
$\rho(B)$ refers to the spectral radius of the matrix $B$. 
Then (\ref{edde}) is satisfied by choosing 
\[ A=\fl{1}{k}\left(\left(I-k B\right)^{-1}-I\right) = 
\left(I-k B\right)^{-1} B.\]
\end{remark}

\begin{remark}\label{fors}
If a discretization for the normalized 
gradient flow (\ref{ngfsp1})-(\ref{ngfsp3})  
reads
\be
\label{dgeff}
\tilde {U}^{n+1}= B U^{n}, 
\qquad U^{n+1}=\fl{\tilde {U}^{n+1}}{\|\tilde {U}^{n+1}\|}, 
\qquad n=0,1,2,\cdots.
\ee
Suppose $B$ is symmetric and positive definite and $\rho(B)<1$. 
Then (\ref{edde}) is satisfied by choosing 
\[ A=\fl{1}{k}\left(B^{-1}-I\right).\]
\end{remark}

\begin{remark}\label{forcn}
If a discretization for the normalized 
gradient flow (\ref{ngfsp1})-(\ref{ngfsp3})  
reads
\be
\label{dgefcn}
\fl{\tilde {U}^{n+1}-U^n}{k} = -B \tilde{U}^{n+1}-C U^n, 
\quad U^{n+1}=\fl{\tilde {U}^{n+1}}{\|\tilde {U}^{n+1}\|}, 
\quad n=0,1,2,\cdots.
\ee
Suppose $B$ and $C$ are symmetric,
positive definite and $\rho(kC)<1$. 
Then (\ref{edde}) is satisfied by choosing 
\[ A=\left(I-k C\right)^{-1}(B+C). \]
\end{remark}


\section{Numerical methods and energy diminishing} 
\label{smethod} 
\setcounter{equation}{0} 
 
In this section, we will present two numerical methods to discretize
the normalized gradient flow (\ref{ngf1})-(\ref{ngf3}).
 For simplicity of 
notation we shall introduce the methods for the case of one spatial 
dimension $(d=1)$ with 
homogeneous periodic boundary conditions.
 Generalizations to $d>1$ are straightforward 
for tensor product grids and the results remain valid without 
modifications. For $d=1$, the problem becomes 
\begin{eqnarray} \label{sdge1d} 
&&\phi_t = \fl{1}{2}\phi_{xx} - V(x) \phi -\bt\; |\phi|^2\phi, \quad 
x\in\Og=(a,b),\ t_n<t<t_{n+1}, \ n\ge0,\qquad \\
\label{sdge1d2}
&&\phi(x,t_{n+1})\stackrel{\triangle}{=}
\phi(x,t_{n+1}^+)=\fl{\phi(x,t_{n+1}^-)}{\|\phi(\cdot,t_{n+1}^-)\|}, 
\qquad a\le x\le b, \quad n\ge 0,\\
\label{sdge1d3}
&&\phi(x,0)=\phi_0(x), \qquad a\le x \le b,\\
\label{sdge1d4}
&&\phi(a,t)=\phi(b,t)=0, \qquad t\ge0;
\eea
with 
\[\|\phi_0\|^2=\int_a^b \phi_0^2(x)\; dx=1.\]

\subsection{Numerical methods}
We choose the spatial mesh size $h=\btu x>0$ with $h=(b-a)/M$ and 
$M$ an even positive integer, the time step is given by $k=\btu 
t>0$ and define grid points and time steps by 
\[ 
x_j:=a+j\;h, \qquad t_n := n\; k, \qquad j=0,1,\cdots, M, \qquad 
n=0,1,2,\cdots 
\] 
Let $\phi^{n}_j$ be the numerical approximation of $\phi(x_j,t_n)$ 
and $\phi^{n}$ the solution vector at time $t=t_n=nk$ with 
components $\phi_j^{n}$. 
 
\bigskip

{\bf Backward Euler finite difference (BEFD)} We use backward Euler
for time discretization and second-order centered finite difference
for spatial derivatives. The detail scheme is:
\bea
&&\fl{\phi_j^*-\phi_j^n}{k}=\fl{1}{2h^2}\left[\phi_{j+1}^*
-2\phi_j^*+\phi_{j-1}^*\right]-V(x_j)\phi_j^*-\bt \left(\phi_j^n\right)^2
\phi_j^*,  \quad j=1,\cdots, M-1,\nn\\
&&\phi_0^*=\phi_M^*=0,\nn\\
\label{befd3}
&&\phi_j^{n+1}=\fl{\phi_j^*}{\|\phi^*\|}, \qquad j=0,\cdots, M, 
\qquad n=0,1,\cdots,\\
&&\phi_j^0= \phi_0(x_j), \qquad j=0,1,\cdots, M; \nn
\eea 
where the norm is defined as
\[\|\phi^*\|^2 = h\sum_{j=1}^{M-1} \left(\phi_j^*\right)^2.\]

\bigskip

{\bf Time-splitting sine-spectral method (TSSP)} 
From time $t=t_n$ to time $t=t_{n+1}$, the equation
(\ref{sdge1d}) is solved in two steps. 
One solves 
\begin{equation} \label{fstep} 
\phi_t=\fl{1}{2} \phi_{xx}, 
\end{equation} 
for one time step of length $k$, followed by solving 
\begin{equation} 
\label{sstep} 
\phi_t(x,t)= -V(x)\phi(x,t)- \bt |\phi|^2\phi(x,t), \qquad t_n\le t\le t_{n+1},
\end{equation} 
again for the same time step. Equation (\ref{fstep}) is 
discretized in space by the sine-spectral method and integrated in 
time {\it exactly}. For $t\in[t_n,t_{n+1}]$, multiplying the ODE 
(\ref{sstep}) by $\phi(x,t)$, one obtains with $\rho(x,t)=\phi^2(x,t)$ 
\begin{equation} \label{sstepa} 
\rho_t(x,t)= -2 V(x)\rho(x,t) -2\bt \rho^2(x,t), \qquad t_n\le t\le t_{n+1}.
\end{equation} 
The solution of the ODE (\ref{sstepa}) can be expressed as
\be
\label{srho}
\rho(x,t)=\left\{\ba{ll}
\dpm\fl{V(x)\rho(x,t_n)}{\left(V(x)+\bt \rho(x,t_n)\right)e^{2V(x)(t-t_n)}
-\bt \rho(x,t_n)} &\qquad V(x)\ne 0,\\
\ \\ 
\dpm\fl{\rho(x,t_n)}{1+2\bt \rho(x,t_n)(t-t_n)}, &\qquad V(x)=0.
\ea\right.
\ee
Combining the splitting step via the standard second-order Strang splitting 
for solving the normalized gradient flow (\ref{sdge1d})-(\ref{sdge1d4}), 
in detail, the steps for obtaining $\phi_j^{n+1}$ from 
$\phi_j^n$ are given by
\begin{eqnarray} 
&&\phi^*_j=\left\{\ba{ll}
\dpm\sqrt{\fl{V(x_j)e^{-k V(x_j)}}{V(x_j)+\bt (1-e^{-k V(x_j)})|\phi_j^n|^2}}
\ \phi_j^n &\qquad V(x_j)\ne 0,\\
\ \\ 
\dpm\fl{1}{\sqrt{1+\bt k |\phi_j^n|^2}}\ \phi_j^n, &\qquad V(x_j)=0,
\ea\right. \nn\\
&&\phi_j^{**}=\sum_{l=1}^{M-1} e^{-k\mu_l^2/2}\;\widehat{\phi}^*_l \; 
 \sin(\mu_l(x_j-a)),\qquad j=1,2,\cdots,M-1,\nn\\ 
\label{schmg} 
&&\phi^{***}_j=\left\{\ba{ll}
\dpm\sqrt{\fl{V(x_j)e^{-k V(x_j)}}{V(x_j)+\bt (1-e^{-k V(x_j)})
|\phi_j^{**}|^2}} \ \phi_j^{**} &\qquad V(x_j)\ne 0,\\
\ \\ 
\dpm\fl{1}{\sqrt{1+\bt k |\phi_j^{**}|^2}}\ \phi_j^{**}, &\qquad V(x_j)=0,
\ea\right. \nn\\
&&\phi_j^{n+1}=\fl{\phi_j^{***}}{\|\phi^{***}\|}, \qquad j=0,\cdots, M, 
\qquad n=0,1,\cdots;
\eea
where $ \widehat{U}_l$ are the sine-transform coefficients of a 
real vector $U=(u_0,u_1, \cdots, u_M)^T$ with $u_0=u_M=0$ 
which are defined as 
\begin{equation} 
\label{Fourc1} \mu_l=\fl{\pi l}{b-a}, \quad \widehat{U}_l= 
\fl{2}{M}\sum_{j=1}^{M-1} u_j\;\sin(\mu_l (x_j-a)), \quad 
l=1,2,\cdots, M-1
\end{equation} 
and
\[ 
\phi^{0}_j=\phi(x_j,0)=\phi_0(x_j), \qquad j=0,1,2,\cdots,M. 
\] 
Note that the only time discretization error of TSSP is the 
splitting error, which is second order in $k$.

\bigskip

For comparison purposes we review a few other numerical 
methods which are currently used for solving 
the normalized gradient flow. One is the
Crank-Nicolson finite difference (CNFD) scheme \cite{Tomio}:
\bea
&&\fl{\phi_j^*-\phi_j^n}{k}=\fl{1}{4h^2}\left[\phi_{j+1}^*
-2\phi_j^*+\phi_{j-1}^*+\phi_{j+1}^n
-2\phi_j^n+\phi_{j-1}^n\right]\nn\\
&&\qquad \qquad -\fl{V(x_j)}{2}\left[\phi_j^*+\phi_j^n\right]
-\fl{\bt \left|\phi_j^n\right|^2}{2}\left[\phi_j^*+\phi_j^n\right],  
\qquad j=1,\cdots, M-1,\nn\\
&&\phi_0^*=\phi_M^*=0,\nn\\
\label{cnfd3}
&&\phi_j^{n+1}=\fl{\phi_j^*}{\|\phi^*\|}, \qquad j=0,\cdots, M, 
\qquad n=0,1,\cdots,\\
&&\phi_j^0= \phi_0(x_j), \qquad j=0,1,\cdots, M. \nn
\eea 
Another one is the forward Euler finite difference 
(FEFD) method \cite{Tosi}:
\bea
&&\fl{\phi_j^*-\phi_j^n}{k}=\fl{1}{2h^2}\left[\phi_{j+1}^n
-2\phi_j^n+\phi_{j-1}^n\right]-V(x_j)\phi_j^n-\bt \left|\phi_j^n\right|^2
\phi_j^n ,  
\qquad j=1,\cdots, M-1,\nn\\
&&\phi_0^*=\phi_M^*=0,\nn\\
\label{fefd3}
&&\phi_j^{n+1}=\fl{\phi_j^*}{\|\phi^*\|}, \qquad j=0,\cdots, M, 
\qquad n=0,1,\cdots,\\
&&\phi_j^0= \phi_0(x_j), \qquad j=0,1,\cdots, M; \nn
\eea 

\subsection{Energy diminishing}

  First we analyze the energy diminishing of the different numerical methods
for linear case, i.e. $\bt=0$ in (\ref{sdge1d}). Introducing
\beas
&&\Phi^n=\left(\phi_1^n,\;\phi_2^n,\;\cdots,\;\phi_{M-1}^n\right)^T,\\
&&D=\left(d_{jl}\right)_{(M-1)\tm (M-1)}, \quad \hbox{with}
\ d_{jl}=\fl{1}{2h^2}\left\{\ba{ll}
2 &j=l,\\
-1 &|j-l|=-1,\\
0 &\hbox{otherwise},
\ea\right. \quad j,l=1,\cdots,M-1,\\
&&E={\rm diag}\left(V(x_1),\; V(x_2),\; \cdots,\; V(x_{M-1})\right),\\
&&F(\Phi)={\rm diag}\left(\phi_1^2,\;\phi_2^2,\;\cdots, \; 
\phi_{M-1}^2\right),
\qquad \hbox{with}\quad \Phi=\left(\phi_1,\;\phi_2,\;\cdots,
\;\phi_{M-1}\right)^T,\\
&&G=\left(g_{jl}\right)_{(M-1)\tm (M-1)}, \qquad \hbox{with}\ 
g_{jl}=\fl{2}{M}\sum_{m=1}^{M-1}\sin\fl{\pi mj}{M}\; \sin\fl{\pi m l}{M}\;
e^{-k \mu_m^2/2}, \\
&&H={\rm diag}\left(e^{-k V(x_1)/2}, \; e^{-k V(x_2)/2},\;
\cdots,\;  e^{-k V(x_{M-1})/2}\right).
\eeas
Then the BEFD discretization (\ref{befd3}) (called BEFD normalized flow) 
with $\bt=0$ can be expressed as
\be
\label{befdvp}
\fl{\Phi^*-\Phi^n}{k} = - (D+E) \Phi^*,  \quad
\Phi^{n+1}= \fl{\Phi^*}{\|\Phi^*\|}, \qquad n=0,1,\cdots.
\ee
The TSSP discretization (\ref{schmg}) (called TSSP normalized flow)
with $\bt=0$ can be expressed as
\be
\label{tsspvp}
\Phi^{***}=H \Phi^{**}= HG \Phi^{*}=HGH\Phi^{n}, \qquad
\Phi^{n+1}= \fl{\Phi^*}{\|\Phi^*\|}, \qquad n=0,1,\cdots.
\ee
The CNFD discretization (\ref{cnfd3}) (called CNFD normalized flow) 
with $\bt=0$ can be expressed as
\be
\label{cnfdvp}
\fl{\Phi^*-\Phi^n}{k} = - \fl{1}{2}(D+E) \Phi^*-\fl{1}{2}(D+E) \Phi^n,  
\quad \Phi^{n+1}= \fl{\Phi^*}{\|\Phi^*\|}, \qquad n=0,1,\cdots.
\ee
The FEFD discretization (\ref{fefd3}) (called FEFD normalized flow) 
with $\bt=0$ can be expressed as
\be
\label{fefdvp}
\fl{\Phi^*-\Phi^n}{k} = - (D+E) \Phi^n,  \quad
\Phi^{n+1}= \fl{\Phi^*}{\|\Phi^*\|}, \qquad n=0,1,\cdots.
\ee

  It is easy to see that $D$ and $G$ are symmetric positive definite
matrices. Furthermore $D$ is also an $M$-matrix and
$\rho(D)=\left(1+\cos\fl{\pi}{M}\right)/h^2<2/h^2$ 
and $\rho(G)=e^{-k \mu_1^2/2}<1$. Applying Theorem
\ref{deeg} and Remarks \ref{forw},\ref{fors}\&\ref{forcn}, we have

\begin{theorem}\label{eddall}
Suppose $V(x)\ge0$ and $\bt=0$. We have that

  (i). The BEFD normalized flow (\ref{befd3})
is energy diminishing and monotone
for any $k>0$.

  (ii). The TSSP normalized flow (\ref{schmg}) is energy diminishing for
any $k>0$.

  (iii). The CNFD normalized flow (\ref{cnfd3}) is energy diminishing and
monotone provided that 
\be
\label{condkcn}
k\le \fl{2}{2/h^2+\max_j V(x_j)} = \fl{2h^2}{2+h^2\;\max_j V(x_j)}.
\ee

(iv). The FEFD normalized flow (\ref{fefd3}) is energy diminishing and
monotone provided that 
\be
\label{condkfe}
k\le \fl{1}{2/h^2+\max_j V(x_j)} = \fl{h^2}{2+h^2\;\max_j V(x_j)}.
\ee
\end{theorem}

  For nonlinear case, i.e. $\bt>0$, we only analyze the 
{\em energy} between two steps of 
BEFD flow (\ref{befd3}). In this case, consider
\be
\label{mbefd}
\fl{\tilde \Phi^{n+1}-\Phi^n}{k} = - \left(D+E+\bt F(\Phi^n)\right) 
\tilde \Phi^{n+1}, \qquad \Phi^{n+1}=\fl{\tilde \Phi^{n+1}}
{\|\tilde \Phi^{n+1}\|}.
\ee

\begin{lemma} \label{ngrbefd}
Suppose $V(x)\ge0$, $\bt>0$ and
$\|\Phi^n\|=1$. Then for the flow (\ref{mbefd}), we have 
\be
\label{engnl}
\Etb\left(\tilde \Phi^{n+1}\right) 
\le \Etb\left(\Phi^n\right), \qquad 
\tilde {E}_{\Phi^n}\left(\Phi^{n+1}\right)\le 
\tilde {E}_{\Phi^n}\left(\Phi^{n}\right)
\ee
where
\bea
\label{defE}
\Etb\left(\Phi\right)&=&(\Phi,(D+E+\bt F(\Phi))\Phi)
=\Phi^T(D+E)\Phi+\bt \sum_{j=1}^{M-1}\phi_j^4,\\
\label{defEd}
\tilde {E}_{\Phi^n}\left(\Phi\right)&=&(\Phi,(D+E+\bt F(\Phi^n))\Phi)
=\Phi^T(D+E)\Phi+\bt \sum_{j=1}^{M-1}\phi_j^2\; \left(\phi^n_j\right)^2\; .
\qquad 
\eea
\end{lemma}

\noindent {\bf Proof:} \ Combining (\ref{mbefd}), (\ref{dge}) and
Theorem \ref{deeg}, we have
\bea
\label{imdP}
&&\left(\tilde\Phi^{n+1},\; 
(D+E+\bt F(\Phi^n))\tilde\Phi^{n+1}\right)
\le\fl{\left(\tilde\Phi^{n+1},\; 
(D+E+\bt F(\Phi^n))\tilde\Phi^{n+1}\right)}{\left(\tilde\Phi^{n+1},
\tilde\Phi^{n+1}\right)} \nn \\
&\quad &\qquad \le\fl{\left(\Phi^n,\;
 (D+E+\bt F(\Phi^n))\Phi^n\right)}{
\left(\Phi^n,\Phi^n\right)}=\Etb\left(\Phi^n\right).
\eea
Similar to the proof of (\ref{dphi2}), we have
\be
\label{phi4}
\sum_{j=1}^{M-1} \left(\phi_j^n\right)^2 
\left(\tilde\phi_j^{n+1}\right)^2 
\ge\sum_{j=1}^{M-1}\left(\tilde\phi_j^{n+1}\right)^4.
\ee
The required result (\ref{engnl}) is a combination of
(\ref{phi4}), and (\ref{imdP}). \hfill $\Box$



\section{Numerical results} \label{sne} 
\setcounter{equation}{0} 
\setcounter{figure}{0}

In this section we compare the four different numerical discretizations
for normalized gradient flow and report numerical results
of the ground state solutions of BEC in 1d, 2d and 3d 
with magnetic trap confinement potential. We also compute the
ground state solutions with the potential of a stirrer
corresponding a far-blue detuned Gaussian laser beam 
and central vortex state by the methods BEFD or TSSP.

 Due to the ground state solution $\phi_g(\bx)\ge0$ for $\bx\in\Og$
in non-rotating BEC \cite{Lieb}, in our computations, the initial condition 
(\ref{ngf3}) is always chosen such that $\phi_0(\bx)\ge0$ and
 decays to zero sufficiently fast as $|{\bf x}|\to\ift$. We choose an 
appropriately large interval, rectangle and box in 1d, 2d and 3d, 
respectively, to avoid that 
the homogeneous periodic boundary condition (\ref{sdge1d4}) 
introduce a significant (aliasing) error relative to the whole 
space problem. To quantify the ground state solution $\phi_g(\bx)$,
we define the radius mean square 
\be
\label{rms}
\ap_{\rm rms}=\|\ap \phi_g\|_{L^2(\Og)}=
\sqrt{\int_\Og \ap^2 \phi_g^2(\bx)\;d\bx}, 
\qquad \ap=x,y,\ {\rm or} \ z.
\ee

\subsection{Comparisons of different methods}

{\bf Example 1} Normalized gradient flow in 1d, i.e.
$d=1$ in (\ref{ngf1})-(\ref{ngf3}). We consider
two cases:

\noindent {\sl I. Linear case ($\bt=0$) 
with a double-well potential}, 
\[V(x)=\fl{1}{2}(1-x^2)^2, \qquad \bt=0, 
\quad \phi_0(x)=\fl{1}{(4\pi)^{1/4}}e^{-x^2/8}, \quad x\in {\mathbb R}.\]

\noindent {\sl II. Nonlinear case ($\bt>0$) with 
a harmonic oscillator potential},  
\[V(x)=\fl{x^2}{2}, \qquad \bt=60, 
\quad \phi_0(x)=\fl{1}{(\pi)^{1/4}}e^{-x^2/2}, \quad x\in {\mathbb R}.\]

The case I is solved on $\Og=[-16, 16]$ and the case II on $\Og=[-8,8]$ 
with mesh size $h=\fl{1}{32}$. 
Figure 4.1 shows the evolution of the
energy $\Eb(\phi)$ for different time step $k$ and 
different numerical methods.

\bigskip

From Figure 4.1, 
the following observations can be made:

(1). BEFD is an implicit method and energy diminishing is observed
for both linear and the nonlinear case under any time step $k>0$. 
The error in the ground state solution
is only due to the second order spatial discretization.

 
(2). TSSP is an explicit method and energy diminishing is observed
for linear case under any time step $k>0$. For nonlinear case,
our numerical experiments show that $k<\fl{1}{\bt}$ guarantees
energy diminishing. The error in the ground state solution
is caused by both the spatial discretization which is spectral accuracy
and time splitting which is second-order accuracy. 
From accuracy point of view, 
large values of $k$ should be prohibited.

(3). CNFD is an implicit method and FEFD is an explicit method.
For both schemes, energy diminishing is observed only when the time step 
$k$ satisfies the condition (\ref{condkcn}) and
(\ref{condkfe}), respectively. 

\bigskip

To summarize briefly, in general, BEFD is much better than 
CNFD for computing the
ground state solution because BEFD is monotone for any $k>0$ 
and CNFD is {\bf not}. TSSP is much better than FEFD. In practice,
one can use either BEFD or TSSP. BEFD allows the use of
much bigger time step $k$ 
which does {\bf not} depend on $\bt\ge0$, but the scheme has only
second order accuracy in space. At each time step, a linear system
is solved. 
In the appendix, 
we give the detailed BEFD discretization in 2d and 3d when 
the potential $V(\bx)$ and the initial data 
$\phi_0(\bx)$ have symmetry with/without a
central vortex state in the condensate.  
TSSP is explicit, easy to program, less demanding on memory 
 and spectrally accurate
in space,  but it needs a small time step $k$
which depends on the accuracy required and the value of $\bt>0$,
 but not on  the mesh size $h$.
Based on our numerical experiments given in the next subsection, 
both methods work very well for computing the ground state 
solution of BEC.

\subsection{Applications to ground state solutions}

{\bf Example 2} Ground state solution of BEC in 1d with harmonic oscillator
potential, i.e. 
\[V(x)=\fl{x^2}{2}, 
\qquad \phi_0(x)=\fl{1}{(\pi)^{1/4}}e^{-x^2/2}, \quad x\in {\mathbb R}.\]
The normalized gradient flow 
(\ref{ngf1})-(\ref{ngf3}) with $d=1$ is solved on 
$\Og=[-16,16]$ with mesh size $h=\fl{1}{8}$ and
time step $k=0.001$ by using TSSP. The steady state solution
is reached when $\max\left|\Phi^{n+1}-\Phi^n\right|<\vep=10^{-6}$.
Figure 4.2 
shows the ground state solution $\phi_g(x)$ 
and energy evolution
for different $\bt$.  Table \ref{tab1t} displays the values of
 $\phi_g(0)$, radius mean square $x_{\rm rms}$, energy $\Eb(\phi_g)$ and 
chemical potential $\mu_g$.

\begin{table}[htbp] \label{tab1t}
\begin{center}
\caption{Maximum value of the wave function $\phi_g(0)$, root mean square 
size $x_{\rm rms}$, energy
$\Eb(\phi_g)$ and ground state 
chemical potential $\mu_g$ versus the interaction coefficient 
$\bt$ in 1d.}
\begin{tabular}{ccccc}\hline
$\bt$  &$\phi_g(0)$ &$x_{\rm rms}$  &$\Eb(\phi_g)$ &$\mu_g=\mu_\bt(\phi_g)$ 
  \\ \hline
  0    &0.7511       &0.7071     &0.5000      &0.5000\\
  3.1371  &0.6463       &0.8949     &1.0441      &1.5272\\ 
 12.5484   &0.5301     &1.2435     &2.2330      &3.5986\\
 31.371   &0.4562    &1.6378     &3.9810      &6.5587\\ 
 62.742  &0.4067  &2.0423     &6.2570      &10.384\\
156.855  &0.3487  &2.7630     &11.464      &19.083\\  
313.71      &0.3107  &3.4764     &18.171      &30.279\\ 
627.42      &0.2768  &4.3757   &28.825      &48.063\\    
1254.8   &0.2467   &5.5073     &45.743      &76.312\\   
   \hline
\end{tabular}
\end{center}
\end{table}

   The results in Figure 4.2 
and Table \ref{tab1t}
 agree very well with the 
ground state solutions of BEC obtained by a direct minimization the
energy functional \cite{Bao}. BEFD gives the same results with $k=0.1$.

  {\bf Example 3} Ground state solution of BEC in 2d.
Two  cases are considered:
\bigskip

{\sl I. With a harmonic oscillator potential \cite{Bao,Bao3,Edwards}}, i.e.
\[V(x,y)=\fl{1}{2}\left(\gm_x^2 x^2+\gm_y^2 y^2\right).\]

{\sl II. With a harmonic oscillator potential and a potential
 of a stirrer
corresponding a far-blue detuned Gaussian laser beam \cite{Jackson}
which is used to generate vortices in BEC \cite{CD1}}, i.e.
\[V(x,y)=\fl{1}{2}\left(\gm_x^2 x^2+\gm_y^2 y^2\right)+w_0
e^{-\dt((x-r_0)^2+y^2)}.\]
The initial condition is chosen as
\[\phi_0(x,y)=\fl{(\gm_x \gm_y)^{1/4}}{\pi^{1/2}}e^{-(\gm_x x^2+\gm_y 
y^2)/2}.\]

For case I, we choose $\gm_x=1$, $\gm_y=4$, $w_0=\dt=r_0=0$, $\bt=200$
and solve the problem by TSSP on $\Og=[-8,8]\tm[-4,4]$ 
with mesh size $h_x=\fl{1}{8}$,
$h_y=\fl{1}{16}$
and time step $k=0.001$.
We get the following results from the
ground state solution $\phi_g$: 
\[x_{\rm rms}=2.2734, \ y_{\rm rms}=0.6074, \
\phi_g^2(0)=0.0808, \  \Eb(\phi_g)=11.1563, \
\mu_g=16.3377.\]

For case II, we choose $\gm_x=1$, $\gm_y=1$, $w_0=4$, $\dt=r_0=1$, $\bt=200$
and solve the problem by TSSP 
on $\Og=[-8,8]^2$ with mesh size $h=\fl{1}{8}$
and time step $k=0.001$. 
We get the following results from the
ground state solution $\phi_g$: 
\[x_{\rm rms}=1.6951, \ y_{\rm rms}=1.7144, \ 
\phi_g^2(0)=0.034, \ \Eb(\phi_g)=5.8507, \
\mu_g=8.3269.\]
In addition, Figure 4.3 
shows surface plots of the ground state solution
$\phi_g$. BEFD gives similar results with $k=0.1$.

\bigskip

   {\bf Example 4} Ground state solution of BEC in 3d. 
Two  cases are considered:
\bigskip

{\sl I. With a harmonic oscillator potential \cite{Bao,Bao3,Edwards}}, i.e.
\[V(x,y,z)=\fl{1}{2}\left(\gm_x^2 x^2+\gm_y^2 y^2+\gm_z^2 z^2\right).\]

{\sl II. With a harmonic oscillator potential and a potential
 of a stirrer
corresponding a far-blue detuned Gaussian laser beam \cite{Jackson,CD2}
which is used to generate vortex in BEC \cite{CD2}}, i.e.
\[V(x,y,z)=\fl{1}{2}\left(\gm_x^2 x^2+\gm_y^2y^2+\gm_z^2 z^2\right)+w_0
e^{-\dt((x-r_0)^2+y^2)}.\]


The initial condition is chosen as
\[\phi_0(x,y,z)=\fl{(\gm_x \gm_y\gm_z)^{1/4}}{\pi^{3/4}}
e^{-(\gm_x x^2+\gm_y y^2+\gm_z z^2)/2}.\]

For case I, we choose $\gm_x=1$, $\gm_y=2$, $\gm_z=4$,
$w_0=\dt=r_0=0$, $\bt=200$
and solve the problem by TSSP on $\Og=[-8,8]\tm[-6,6]\tm[-4,4]$ 
with mesh size $h_x=\fl{1}{8}$, $h_y=\fl{3}{32}$, $h_z=\fl{1}{16}$
and time step $k=0.001$.
The ground state solution $\phi_g$ gives: 
\[x_{\rm rms}=1.67, \ y_{\rm rms}=0.87, \ z_{\rm rms}=0.49,\
\phi_g^2(0)=0.052, \ \Eb(\phi_g)=8.33, \
\mu_g=11.03.\]

For case II, we choose $\gm_x=1$, $\gm_y=1$, $\gm_z=2$, 
$w_0=4$, $\dt=r_0=1$, $\bt=200$
and solve the problem  by TSSP
on $\Og=[-8,8]^3$ with mesh size $h=\fl{1}{8}$
and time step $k=0.001$.
The ground state solution $\phi_g$ gives: 
\[x_{\rm rms}=1.37, \ y_{\rm rms}=1.43, \ z_{\rm rms}=0.70,
\ \phi_g^2(0)=0.025, \ \Eb(\phi_g)=5.27, \
\mu_g=6.71.\]

Furthermore, Figure 4.4 
shows surface plots of the ground 
state solution $\phi_g^2(x,0,z)$. BEFD gives similar results 
with $k=0.1$.

\bigskip

{\bf Example 5} 2d central vortex states in BEC, i.e. 
\[V(x,y)=V(r)=\fl{1}{2}\left(\fl{m^2}{r^2}+r^2\right), 
\ \phi_0(x,y)=\phi_0(r)=\fl{1}{\sqrt{\pi m!}}\;r^m\;e^{-r^2/2},
\ 0\le r.\]
The normalized gradient flow is solved in polar coordinate 
with  $\Og=[0,8]$ with mesh size $h=\fl{1}{64}$ and
time step $k=0.1$ by using BEFD (see detail in Appendix A3).
Figure 4.5a 
shows the ground state solution $\phi_g(r)$ with 
$\bt=200$ 
for different index of the central vortex $m$.  
Table \ref{tab2t} displays the values of
 $\phi_g(0)$, radius mean square $r_{\rm rms}$, energy $\Eb(\phi_g)$ and 
chemical potential $\mu_g$.

\begin{table}[htbp]\label{tab2t}
\begin{center}
\caption{Numerical results for 2d central vortex states in BEC.}
\begin{tabular}{ccccc}\hline
Index $m$  &$\phi_g(0)$ &$r_{\rm rms}$  &$\Eb(\phi_g)$ &$\mu_g=\mu_\bt(\phi_g)$
   \\ \hline
  1      &0.0000     &2.4086     &5.8014     &8.2967 \\
  2      &0.0000     &2.5258     &6.3797     &8.7413 \\
  3      &0.0000     &2.6605     &7.0782     &9.3160 \\
  4      &0.0000     &2.8015     &7.8485     &9.9772 \\
   5      &0.0000     &2.9438     &8.6660    &10.6994 \\
  6      &0.0000     &3.0848     &9.5164    &11.4664\\
    \hline
\end{tabular}
\end{center}
\end{table}

\subsection{Application to compute the first excited  state }

 Suppose the eigenfunctions of the nonlinear eigenvalue problem
(\ref{gss}), (\ref{gss1}) under the constraint (\ref{normgg}) are
\[\pm\phi_g(\bx), \ \pm\phi_1(\bx),\ \pm\phi_2(\bx),\ \cdots,\]
whose energies satisfy
\[\Eb(\phi_g)<\Eb(\phi_1)<\Eb(\phi_2)<\cdots \;.\]
Then $\phi_j$ is called as the $j$-th excited state solution. In fact,
$\phi_g$ and $\phi_j$ ($j=1,2,\cdots$) are critical points 
of the energy functional $E_\bt(\phi)$ under the constraint (\ref{normgg}).
In 1d, when $V(x)=\fl{x^2}{2}$ is chosen as the harmonic oscillator potential,
 the first excited state solution $\phi_1(x)$ is
a real odd function, and $\phi_1(x)=\fl{\sqrt{2}}{(\pi)^{1/4}}\; x\; 
e^{-x^2/2}$ when $\bt=0$ \cite{Levine}.
We observe numerically that the normalized gradient flow 
(\ref{ngf1})-(\ref{ngf3}) and its BEFD 
discretization (\ref{befd3}) can also be applied
directly to compute the first excited state solution, i.e. $\phi_1(x)$,
provided that the initial data $\phi_0(x)$ in (\ref{ngf3})
is chosen as an odd function. Here we only present a preliminary 
numerical example in 1d. Extensions to 2d and 3d are straightforward.

\bigskip

{\bf Example 6} First excited state solution of BEC in 1d 
with a harmonic oscillator potential, i.e. 
\[V(x)=\fl{x^2}{2}, 
\qquad \phi_0(x)=\fl{\sqrt{2}}{(\pi)^{1/4}}\; x\; 
e^{-x^2/2}, \quad x\in {\mathbb R}.\]
The normalized gradient flow 
(\ref{ngf1})-(\ref{ngf3}) with $d=1$ is solved on 
$\Og=[-16,16]$ with mesh size $h=\fl{1}{64}$ and
time step $k=0.1$ by using BEFD. 
Figure 4.5b 
shows the first excited state solution $\phi_1(x)$ 
for different $\bt$.  Table \ref{tab3t} displays the 
 radius mean square $x_{\rm rms}=\|x\phi_1\|_{L^2(\Og)}$, 
ground state and first excited state energies
$\Eb(\phi_g)$ and  $\Eb(\phi_1)$, ratio $\Eb(\phi_1)/\Eb(\phi_g)$,  
chemical potentials $\mu_g=\mu_\bt(\phi_g)$ and 
$\mu_1=\mu_\bt(\phi_1)$, ratio $\mu_1/\mu_g$.

\begin{table}[htbp]\label{tab3t}
\begin{center}
\caption{Numerical results for the first excited state solution 
in 1d in Example 6. }
\begin{tabular}{cccccccc}\hline
$\bt$   &$x_{\rm rms}$ &$\Eb(\phi_g)$   &$\Eb(\phi_1)$  
&$\fl{\Eb(\phi_1)}{\Eb(\phi_g)}$ &$\mu_g$ &$\mu_1$
 &$\fl{\mu_1}{\mu_g}$
 \\ \hline
  0     &1.2247 &0.500 &1.500  &3.000 &0.500   &1.500 &3.000\\
3.1371  &1.3165 &1.044 &1.941  &1.859 &1.527   &2.357 &1.544\\
12.5484 &1.5441 &2.233 &3.037  &1.360 &3.598   &4.344 &1.207 \\
31.371  &1.8642 &3.981 &4.743  &1.192 &6.558   &7.279 &1.110 \\
62.742  &2.2259 &6.257 &6.999  &1.119 &10.38  &11.089 &1.068\\
156.855 &2.8973 &11.46 &12.191 &1.063 &19.08  &19.784 &1.037\\
313.71  &3.5847 &18.17 &18.889 &1.040 &30.28  &30.969  &1.023\\
627.42  &4.4657 &28.82 &29.539 &1.025 &48.06 &48.733 &1.014\\
1254.8  &5.5870 &45.74 &46.453 &1.016 &76.31 &76.933 &1.008 \\
   \hline
\end{tabular}
\end{center}
\end{table}

\bigskip

From the results in Table \ref{tab3t} and Figure \ref{fig6t}b, 
we can see that the BEFD
can be applied directly to compute the first excited states in BEC. 
Furthermore, we have
\[\lim_{\bt\to +\ift}\ \fl{\Eb(\phi_1)}{\Eb(\phi_g)}=1, \qquad
 \lim_{\bt\to +\ift}\ \fl{\mu_1}{\mu_g}=1.\]
These results are confirmed with the results 
in \cite{Bao} where the ground and first excited states are computed 
by directly minimizing the energy functional 
through the finite element discretization.

\section{Conclusions}\label{sc} 
\setcounter{equation}{0} 
 
Energy diminishing of a normalized gradient flow and its discretization
are examined, which provides some mathematical justification of the 
imaginary time integration method 
used in physical literatures to compute the ground state solution 
of Bose-Einstein condensation (BEC). 
Backward Euler centered finite difference (BEFD) and
time-splitting sine-spectral (TSSP) method are
proposed  to discretize the normalized gradient flow.
 Comparison between the two proposed methods and existing
methods shows that BEFD and TSSP are 
much better for the  computation of the BEC ground state solution. 
Numerical results in 1d, 2d and 3d with different
types of potentials used in BEC   are reported
to demonstrate the effectiveness of the BEFD and TSSP methods. 
Furthermore, extension of the normalized 
gradient flow and its BEFD discretization to compute higher excited states
with  an orthonormalization technique is on-going.


\bigskip
\bigskip

\renewcommand{\theequation}{\Alph{section}.\arabic{equation}} 
\begin{center}
{\bf Appendix: BEFD discretization in BEC when $V(\bx)$ has symmetry}
\end{center}

  In this appendix, we present detailed BEFD discretizations 
for the normalized gradient flows in BEC in 2d and 3d when 
the potential $V(\bx)$ and the initial data 
$\phi_0(\bx)$ have symmetry with/without a
central vortex state in the condensate.  
  Choose $R>0$, $a<b$ and time step $k>0$ 
with $|a|$, $b$, $R$ sufficiently large.
Denote the mesh size
$h_r=(R-0)/M$ and $h_z=(b-a)/N$ with $M$ and $N$ two positive integers,
time steps $t_n=n\;k$, $n=0,1,\cdots$,
 and grid points $r_j=j\, h_r$, $j=0,1,\cdots, M$ and
$r_{j-\fl{1}{2}}=\left(j-\fl{1}{2}\right)h_r$, $j=0,1,\cdots, M+1$,
$z_l=a+l\,h_z$, $l=0,1,\cdots,N$.

\bigskip

\noindent {\it A1. 2d with radial symmetry and 3d with spherical symmetry},
i.e. $V(\bx)=V(r)$ and $\phi_0(\bx)=\phi_0(r)$
with $r=|\bx|$ and $\Og={\mathbb R}^d$ with 
$d=2,3$ in (\ref{ngf1})-(\ref{ngf3}). In this case, the solution 
$\phi(\bx,t)=\phi(r,t)$ and the normalized 
gradient flow collapses to a 1d problem:
\bea
\label{2d3d1}
&&\phi_t=\fl{1}{2r^{d-1}}\pl{}{r}\left(r^{d-1}\pl{\phi}{r}\right)
-V(r) \phi -\bt |\phi|^2\phi, \quad 0<r<\ift,\ t_n<t<t_{n+1},
\qquad \\
\label{2d3d2}
&&\phi_r(0,t)=0, \qquad \lim_{r\to \ift} \phi(r,t)=0, \qquad t\ge0,\\
\label{2d3d3}
&&\phi(r,t_{n+1})\stackrel{\triangle}{=}
\fl{\phi(r,t_{n+1}^-)}{\|\phi(\cdot,t_{n+1}^-)\|}, \qquad 
0<r<\ift, \quad n\ge0,\\
\label{2d3d4}
&&\phi(r,0)=\phi_0(r)\ge0, \qquad 0<r<\ift;
\eea
where $\|\phi_0\|=1$ and the norm $\|\cdot\|$ is defined as
\[\|\phi\|^2=C_d \int _0^\ift \phi^2(r,t)r^{d-1}\; dr.\]
with 
\[C_d=\left\{\ba{ll}
2\pi, &\qquad d=2,\\
4\pi, &\qquad d=3.\\
\ea\right. \]
The BEFD discretization of (\ref{2d3d1})-(\ref{2d3d4}) is:
\bea
\label{2d3d1d}
&&\fl{\phi_{j-\fl{1}{2}}^*-\phi_{j-\fl{1}{2}}^n}{k}=
\fl{1}{2\;h_r^2\;r_{j-\fl{1}{2}}^{d-1}}\left[r_j^{d-1}\;
\phi_{j+\fl{1}{2}}^*
 -\left(r_j^{d-1}+r_{j-1}^{d-1}\right)\phi_{j-\fl{1}{2}}^*
+r_{j-1}^{d-1}\;\phi_{j-\fl{3}{2}}^*\right]\nn\\
&&\qquad \qquad -V(r_{j-\fl{1}{2}})\;\phi_{j-\fl{1}{2}}^*-\bt 
  \left(\phi_{j-\fl{1}{2}}^n
\right)^2\phi_{j-\fl{1}{2}}^*,  \quad j=1,\cdots, M-1,\nn\\
&&\phi_{-\fl{1}{2}}^*=\phi_{\fl{1}{2}}^*,\qquad
\phi_{M-\fl{1}{2}}^*=0,\nn\\
&&\phi_{j-\fl{1}{2}}^{n+1}=\fl{\phi_{j-\fl{1}{2}}^*}
{\|\phi^*\|}, \qquad j=0,\cdots, M, 
\qquad n=0,1,\cdots,\\
&&\phi_{j-\fl{1}{2}}^0= \phi_0(r_j), \qquad j=1,\cdots, M,
\qquad \phi_{-\fl{1}{2}}^0=\phi_{\fl{1}{2}}^0, \nn
\eea 
where the norm is defined as
\[\|\phi^*\|^2=h_r C_d \sum_{j=1}^M \left(\phi_{j-\fl{1}{2}}^*\right)^2\; 
r_{j-\fl{1}{2}}^{d-1}. \]

\bigskip

\noindent {\it A2. 3d with cylindrical symmetry}, 
i.e. $V(\bx)=V(r,z)$ and $\phi_0(\bx)=\phi_0(r,z)$
with $r=\sqrt{x^2+y^2}$ and $\Og={\mathbb R}^d$ with 
$d=3$ in (\ref{ngf1})-(\ref{ngf3}). This is the most popular case
in the setup of current BEC experiments.  In this case, the solution 
$\phi(\bx,t)=\phi(r,z,t)$ and the normalized 
gradient flow collapses to a 2d problem with
$0<r<\ift$ and $-\ift<z<\ift$:
\bea
\label{3dcs1}
&&\phi_t=\fl{1}{2}\left[\fl{1}{r}\pl{}{r}
\left(r\pl{\phi}{r}\right)+\fl{\p^2 \phi}{\p z^2}\right]
-V(r,z) \phi -\bt |\phi|^2\phi, 
\quad t_n<t<t_{n+1},\qquad \\
\label{3dcs2}
&&\phi_r(0,z,t)=0, \quad \lim_{r\to \ift} \phi(r,z,t)=0, 
\quad \lim_{z\to \pm \ift}\phi(r,z,t)=0,
\qquad t\ge0,\\
\label{3dcs3}
&&\phi(r,z,t_{n+1})\stackrel{\triangle}{=}
\fl{\phi(r,z,t_{n+1}^-)}{\|\phi(\cdot,t_{n+1}^-)\|}, 
\quad n\ge0,\\
\label{3dcs4}
&&\phi(r,z,0)=\phi_0(r,z)\ge0;
\eea
where $\|\phi_0\|=1$ and the norm $\|\cdot\|$ is defined as
\[\|\phi\|^2=2\pi \int _0^\ift \int_{-\ift}^\ift 
\phi^2(r,z,t)r\; dzdr.\]
The BEFD discretization of (\ref{3dcs1})-(\ref{3dcs4}) is:
\bea
\label{3dcs2d}
&&\fl{\phi_{j-\fl{1}{2}\,l}^*-\phi_{j-\fl{1}{2}\,l}^n}{k}=
\fl{1}{2\;h_r^2\;r_{j-\fl{1}{2}}}\left[r_j\; \phi_{j+\fl{1}{2}\,l}^*
-\left(r_j+r_{j-1}\right)\phi_{j-\fl{1}{2}\,l}^*
+r_{j-1}\;\phi_{j-\fl{3}{2}\,l}^*\right] \nn\\
&&\qquad \qquad +\fl{1}{2h_z^2}\left[\phi_{j-\fl{1}{2}\,l+1}^*
-2\phi_{j-\fl{1}{2}\,l}^*
+\phi_{j-\fl{1}{2}\,l-1}^*\right]
-V(r_{j-\fl{1}{2}},z_l)\;\phi_{j-\fl{1}{2}\,l}^* \nn\\
&&\qquad \qquad -\bt \left(\phi_{j-\fl{1}{2}\,l}^n
\right)^2\;\phi_{j-\fl{1}{2}\,l}^*,  \quad j=1,\cdots, M-1,
\quad l=1,2\cdots, N-1,\nn\\
&&\phi_{-\fl{1}{2}\,l}^*=\phi_{\fl{1}{2}\,l}^*,\quad
\phi_{M-\fl{1}{2}\,l}^*=0, \quad l=1,2,\cdots, N-1, \nn\\
&&\phi_{j-\fl{1}{2}\,0}^* = \phi_{j-\fl{1}{2}\,M}^*=0,
\quad j=0,1,\cdots,M.
\nn\\
&&\phi_{j-\fl{1}{2}\,l}^{n+1}=\fl{\phi_{j-\fl{1}{2}\,l}^*}
{\|\phi^*\|}, \quad j=0,\cdots, M, \quad l=0,1,\cdots,N,
\quad n=0,1,\cdots, \qquad \\
&&\phi_{j-\fl{1}{2}\,l}^0= \phi_0(r_{j-\fl{1}{2}},z_l), 
\qquad j=1,\cdots, M, \quad l=0,\cdots, N, \nn\\
&&\phi_{-\fl{1}{2}\,l}^0=\phi_{\fl{1}{2}\,l}^0, 
\quad  l=0,1,\cdots,N, \nn
\eea 
where the norm is defined as
\[\|\phi^*\|^2=2\pi h_rh_z \sum_{j=1}^M \sum_{l=1}^{N-1}
\left(\phi_{j-\fl{1}{2}\,l}^*\right)^2\; 
r_{j-\fl{1}{2}}. \]

\bigskip

In finding 
a stationary solution of (\ref{sdge}) with a central vortex
state, one plugs the
ansatz 
\[\psi(\bx,t)=\left\{\ba{ll}
e^{-i\mu t}\; e^{i m\tht}\; \phi(r), &\qquad d=2,\\
e^{-i\mu t}\; e^{im\tht}\; \phi(r,z), &\qquad d=3,\\
\ea\right. \qquad r=\sqrt{x^2+y^2}\]
into (\ref{sdge}) instead of (\ref{stat}), where 
$m>0$ an integer corresponding to the index of the vortex. 
For more details 
related to central vortex states in BEC, we refer 
\cite{DalfoS,Konotop,Lundh,Rokhsar}.

\bigskip

\noindent {\it A3. 2d central vortex states in BEC}, 
i.e. $V(\bx)=V(r)=\fl{1}{2}\left(\fl{m^2}{r^2}+r^2\right)$ 
and $\phi_0(\bx)=\phi_0(r) with \phi_0(0)=0$, 
$r=\sqrt{x^2+y^2}$ and $\Og={\mathbb R}^2$  
 in (\ref{ngf1})-(\ref{ngf3}). 
In this case, the solution 
$\phi(\bx,t)=\phi(r,t)$ and the normalized 
gradient flow collapses to a 1d problem:
\bea
\label{2dvs1}
&&\phi_t=\fl{1}{2r}\pl{}{r}\left(r\pl{\phi}{r}\right)
-V(r) \phi -\bt |\phi|^2\phi, \quad 0<r<\ift,\quad t_n<t<t_{n+1},
\qquad \\
\label{2dvs2}
&&\phi(0,t)=0, \qquad \lim_{r\to \ift} \phi(r,t)=0, \qquad t\ge0,\\
\label{2dvs3}
&&\phi(r,t_{n+1})\stackrel{\triangle}{=}
\fl{\phi(r,t_{n+1}^-)}{\|\phi(\cdot,t_{n+1}^-)\|}, \qquad 
0<r<\ift, \quad n\ge0,\\
\label{2dvs4}
&&\phi(r,0)=\phi_0(r)\ge0, \quad 0<r<\ift,  \qquad 
 \left({\rm e.g.} =\fl{1}{\sqrt{\pi m!}}\;r^m\;e^{-r^2/2}\right);
\eea
where $\phi(0)=0$,
 $\|\phi_0\|=1$ and the norm $\|\cdot\|$ is defined as
\[\|\phi\|^2=2\pi \int _0^\ift \phi^2(r,t)r\; dr.\]
The BEFD discretization of (\ref{2dvs1})-(\ref{2dvs4}) is:
\bea
\label{2dvs1d}
&&\fl{\phi_{j}^*-\phi_{j}^n}{k}=
\fl{1}{2\;h_r^2\;r_{j}}\left[r_{j+\fl{1}{2}}\;\phi_{j+1}^*
 -\left(r_{j+\fl{1}{2}}+r_{j-\fl{1}{2}}\right)\phi_{j}^*
+r_{j-\fl{1}{2}}\;\phi_{j-1}^*\right]\nn\\
&&\qquad \qquad -V(r_{j})\;\phi_{j}^*-\bt 
  \left(\phi_{j}^n
\right)^2\phi_{j}^*,  \quad j=1,\cdots, M-1,\nn\\
&&\phi_{0}^*=\phi_{M}^*=0,\nn\\
&&\phi_{j}^{n+1}=\fl{\phi_{j}^*}
{\|\phi^*\|}, \qquad j=0,\cdots, M, 
\qquad n=0,1,\cdots,\\
&&\phi_{j}^0= \phi_0(r_j), \qquad j=0,1,\cdots, M, \nn
\eea 
where the norm is defined as
\[\|\phi^*\|^2=2\pi h_r \sum_{j=1}^{M-1} r_{j}\;
\left(\phi_{j}^*\right)^2. \]

\bigskip

\noindent {\it A4. 3d central vortex states in BEC}, 
i.e. $V(\bx)=V(r,z)=\fl{1}{2}\left(\fl{m^2}{r^2}+\gm_r^2r^2+\gm_z^2
z^2\right)$ and $\phi_0(\bx)=\phi_0(r,z)$ with 
$\phi_0(0,z)=0$ for $z\in{\mathbb R}$, $\gm_r>0$, $\gm_z>0$ constants, 
$r=\sqrt{x^2+y^2}$ and $\Og={\mathbb R}^3$  
 in (\ref{ngf1})-(\ref{ngf3}). 
In this case, the solution 
$\phi(\bx,t)=\phi(r,z,t)$ and the normalized 
gradient flow collapses to a 2d problem with
$0<r<\ift$ and $-\ift<z<\ift$:
\bea
\label{3dvs1}
&&\phi_t=\fl{1}{2}\left[\fl{1}{r}\pl{}{r}
\left(r\pl{\phi}{r}\right)+\fl{\p^2 \phi}{\p z^2}\right]
-V(r,z) \phi -\bt \phi^3, 
\quad t_n<t<t_{n+1},\qquad \\
\label{3dvs2}
&&\phi(0,z,t)=0, \quad \lim_{r\to \ift} \phi(r,z,t)=0, 
\quad \lim_{z\to \pm \ift}\phi(r,z,t)=0,
\qquad t\ge0,\\
\label{3dvs3}
&&\phi(r,z,t_{n+1})\stackrel{\triangle}{=}
\fl{\phi(r,z,t_{n+1}^-)}{\|\phi(\cdot,t_{n+1}^-)\|}, 
\quad n\ge0,\\
\label{3dvs4}
&&\phi(r,z,0)=\phi_0(r,z)\ge0, \qquad 
 \left({\rm e.g.} =\fl{\gm_z^{1/4}\gm_r^{(m+1)/2}}
{\pi^{3/4}(m!)^{1/2}}r^m\;e^{-(\gm_r r^2+\gm_z z^2)/2}\right);
\qquad \quad
\eea
where $\phi_0(0,z)=0$ for $z\in{\mathbb R}$, 
$\|\phi_0\|=1$ and the norm $\|\cdot\|$ is defined as
\[\|\phi\|^2=2\pi \int _0^\ift \int_{-\ift}^\ift 
\phi^2(r,z,t)r\; dzdr.\]
The BEFD discretization of (\ref{3dvs1})-(\ref{3dvs4}) is:
\bea
\label{3dvs2d}
&&\fl{\phi_{j\,l}^*-\phi_{j\,l}^n}{k}=
\fl{1}{2\;h_r^2\;r_{j}}\left[r_{j+\fl{1}{2}}\;\phi_{j+1\,l}^*
-\left(r_{j+\fl{1}{2}}+r_{j-\fl{1}{2}}\right)\phi_{j\,l}^*
+r_{j-\fl{1}{2}}\;\phi_{j-1\,l}^*\right]\nn\\
&&\qquad \qquad +\fl{1}{2h_z^2}\left[\phi_{j\,l+1}^*
-2\phi_{j\,l}^*+\phi_{j\,l-1}^*\right]
-V(r_{j},z_l)\;\phi_{j\,l}^*-\bt \left(\phi_{j\,l}^n
\right)^2\phi_{j\,l}^*, \nn\\
&&\qquad \qquad  \quad j=1,\cdots, M-1,
\ l=1,2\cdots, N-1,\nn\\
&&\phi_{0\,l}^*=\phi_{M\,l}^*=0,\quad l=0,1,\cdots, N, 
\quad \phi_{j\,0}^* = \phi_{j\,M}^*=0,
\quad j=1,1,\cdots,M-1.
\nn\\
&&\phi_{j\,l}^{n+1}=\fl{\phi_{j\,l}^*}
{\|\phi^*\|}, \quad j=0,\cdots, M, \quad l=0,1,\cdots,N,
\quad n=0,1,\cdots, \qquad \\
&&\phi_{j\,l}^0= \phi_0(r_j,z_l), 
\qquad j=0,\cdots, M, \quad l=0,\cdots, N, \nn
\eea 
where the norm is defined as
\[\|\phi^*\|^2=2\pi h_rh_z \sum_{j=1}^{M-1} \sum_{l=1}^{N-1}
\left(\phi_{j\,l}^*\right)^2\; 
r_{j}. \]
The linear system at every time step in A1 and A3 can be solved by
the Thomas algorithm and in A2 and A4 can be solved by Gauss-Seidel 
iterative method.

\setcounter{section}{4}

\begin{figure}[hb] \label{fig2t}
\centerline{a).\psfig{figure=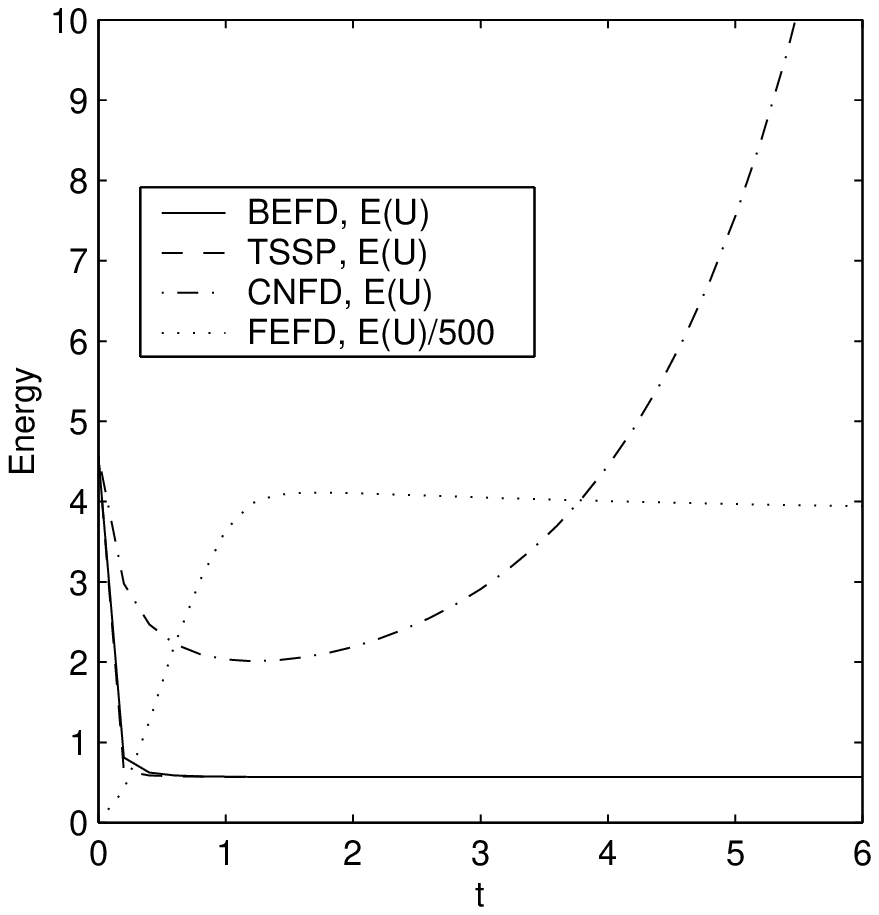,height=6cm,width=6cm,angle=0} 
\qquad b)\psfig{figure=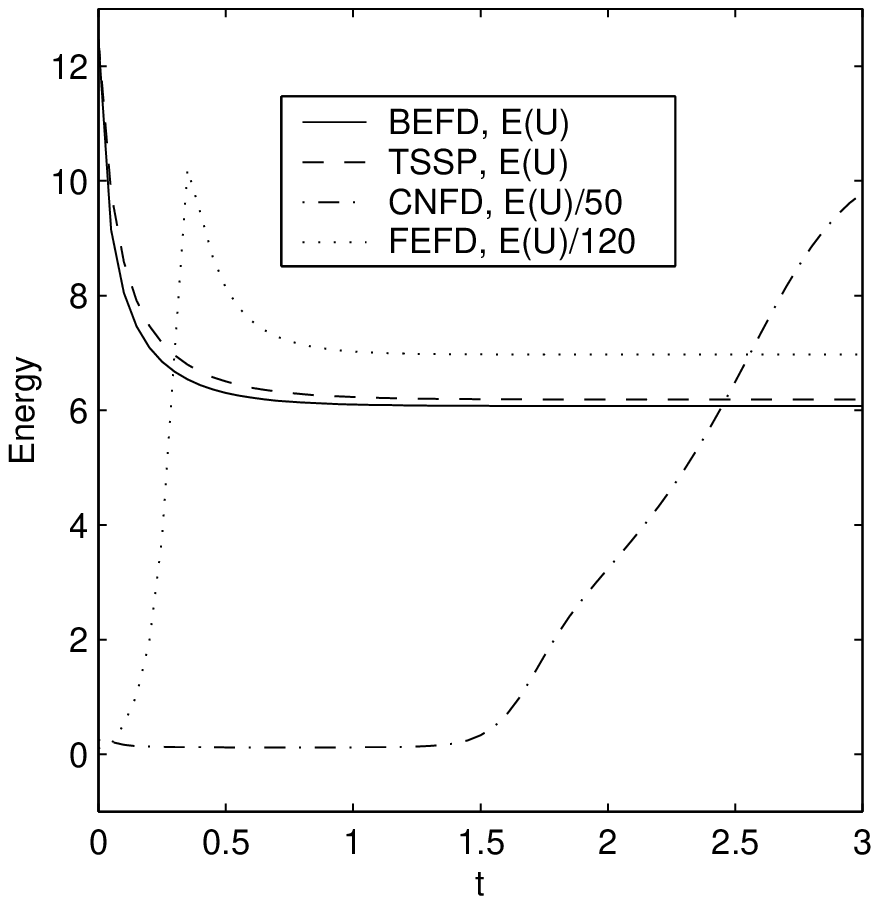,height=6cm,width=6cm,angle=0} } 
\centerline{c).\psfig{figure=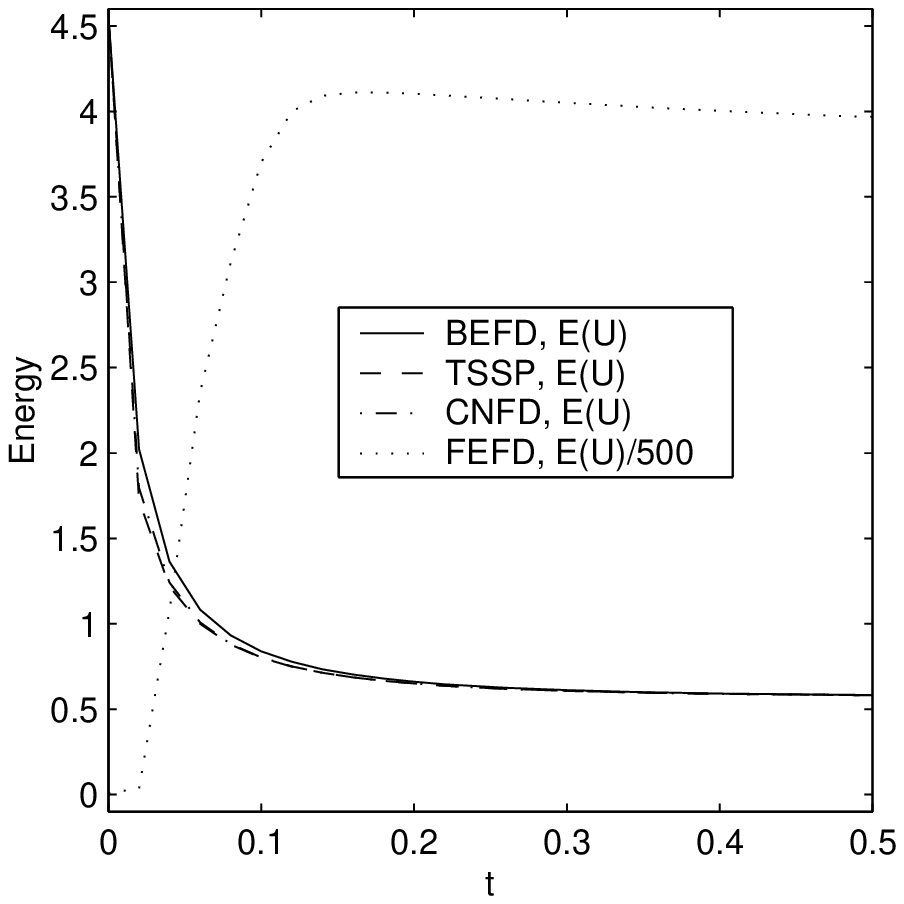,height=6cm,width=6cm,angle=0} 
\qquad d)\psfig{figure=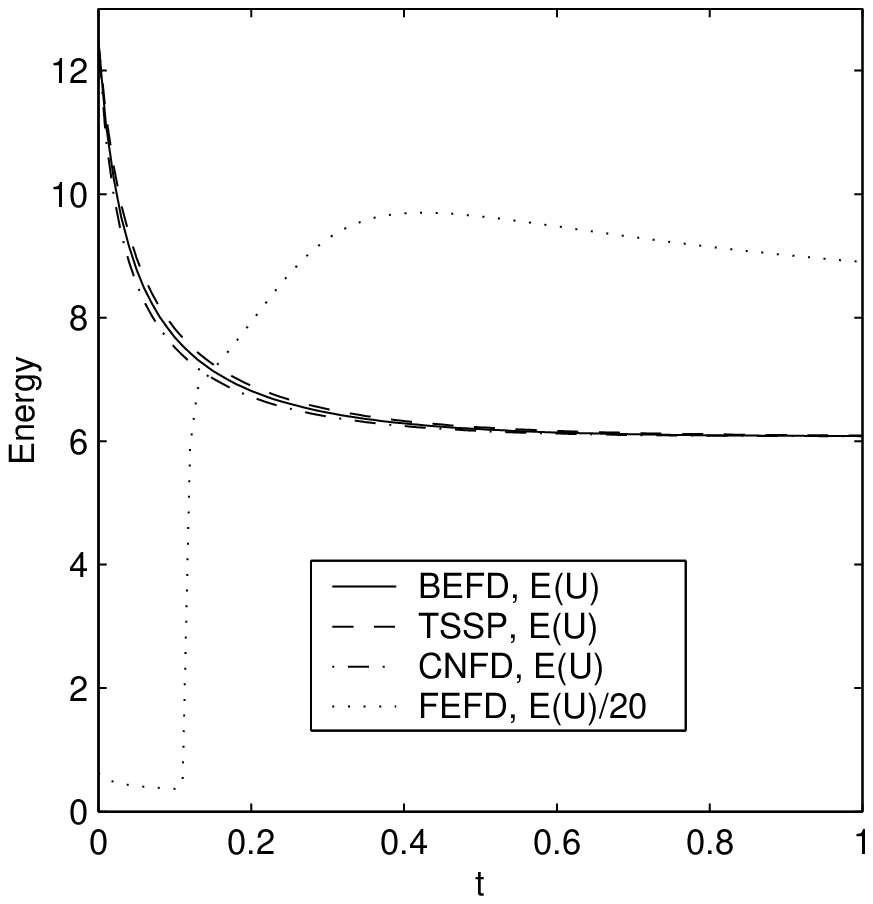,height=6cm,width=6cm,angle=0} } 
\centerline{e).\psfig{figure=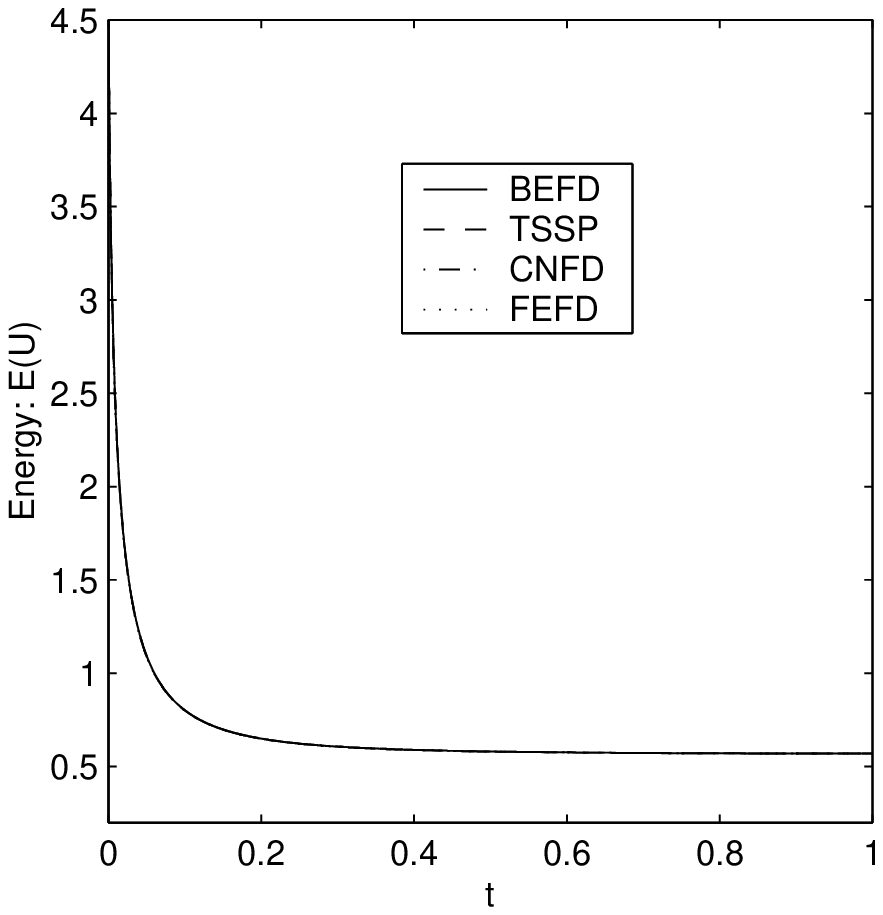,height=6cm,width=6cm,angle=0} 
\qquad f)\psfig{figure=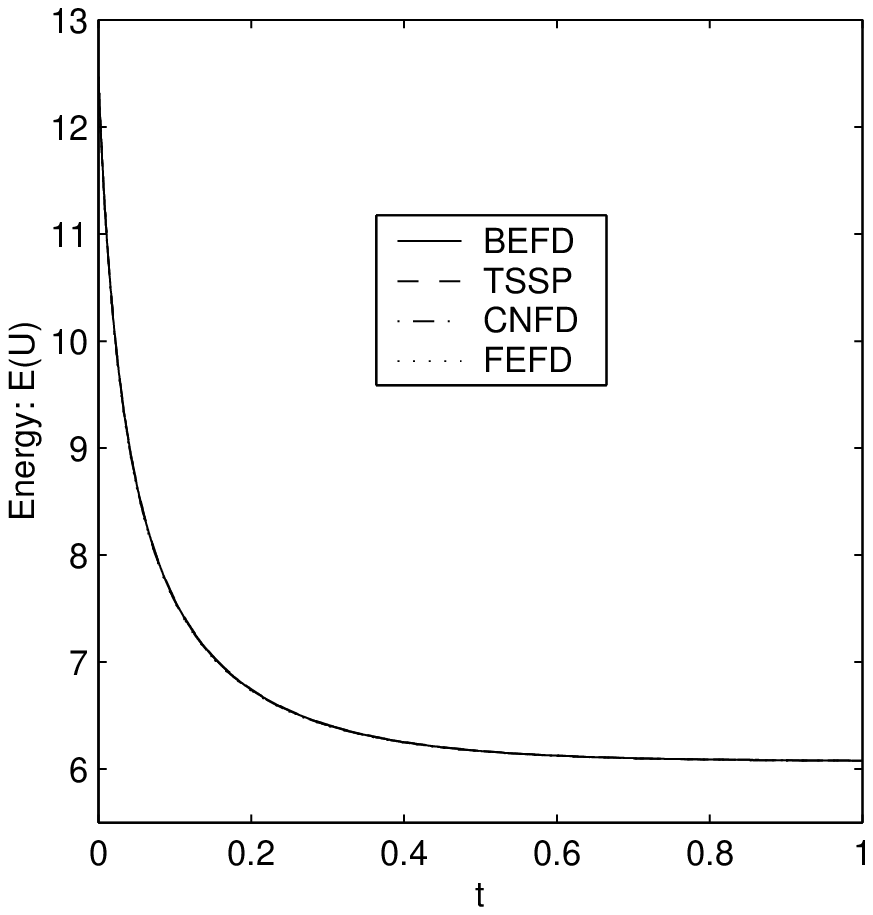,height=6cm,width=6cm,angle=0} } 
\caption{Energy evolution in Example 1. 
Left column for case I: a). $k=0.2$, c). $k=0.02$ and e). $k=0.0005$.
 Right column for case II: b). $k=0.05$, d). $k=0.01$ and
f). $k=0.0005$.}
 \end{figure}

\begin{figure}[hb] \label{fig3t}
\centerline{a).\psfig{figure=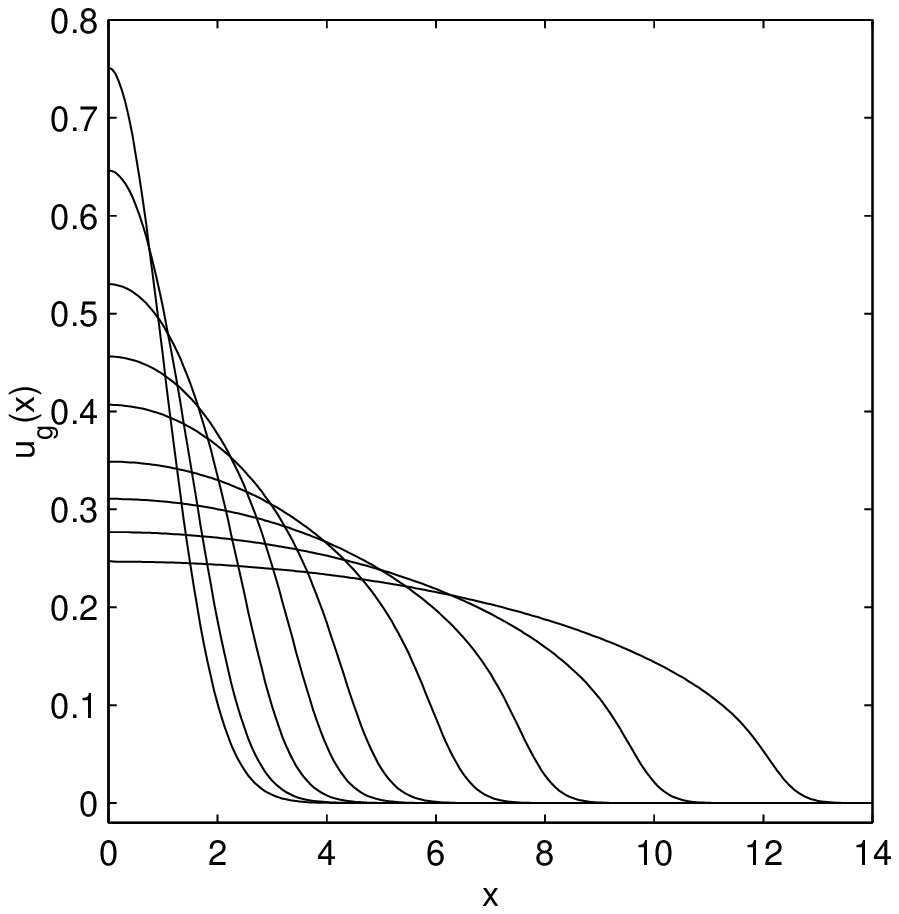,height=6.5cm,width=6cm,angle=0} 
\qquad b).\psfig{figure=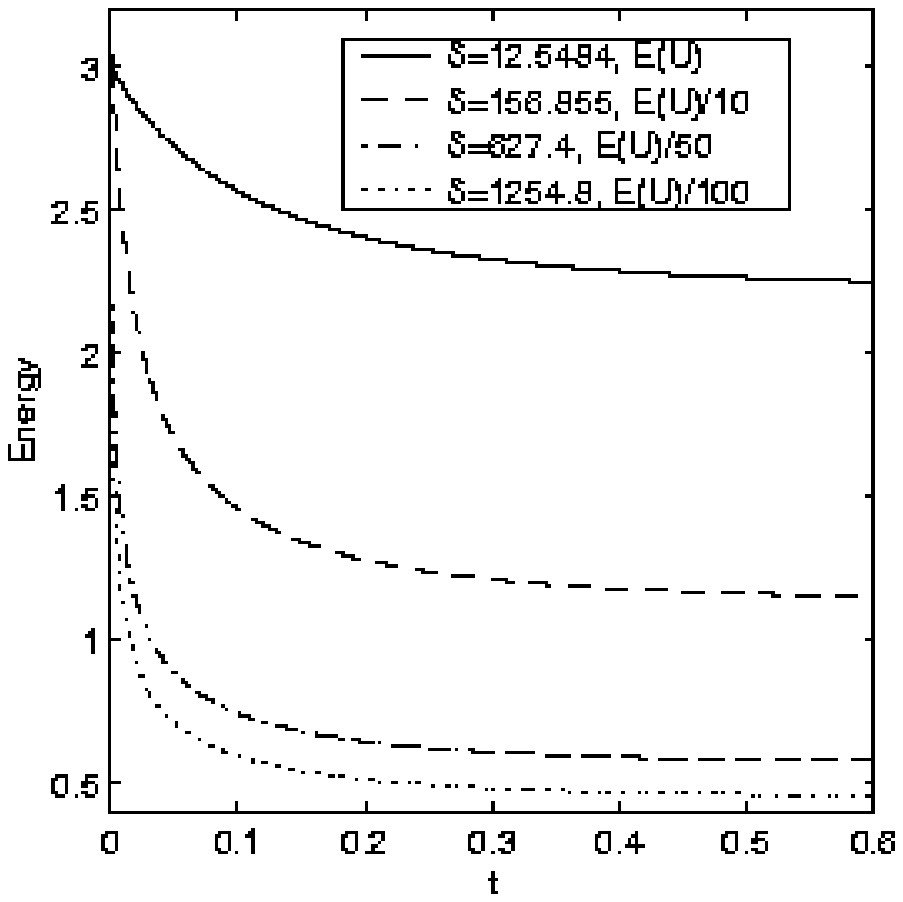,height=6.5cm,width=6cm,angle=0}} 
\caption{Ground state solution $\phi_g(x)$ (labeled as $u_g$)
in Example 2.  (a). For $\bt=0,\; 3.1371,\; 12.5484,\; 31.371,\; 62.742,\; 
156.855,\; 313.71,\; 627.42,\; 1254.8$ (in the order of decreasing peak). 
(b). Energy evolution 
for different $\bt$ (labeled as $\dt$).
}
\end{figure} 

\begin{figure}[hb] \label{fig4t}
\centerline{(a).\psfig{figure=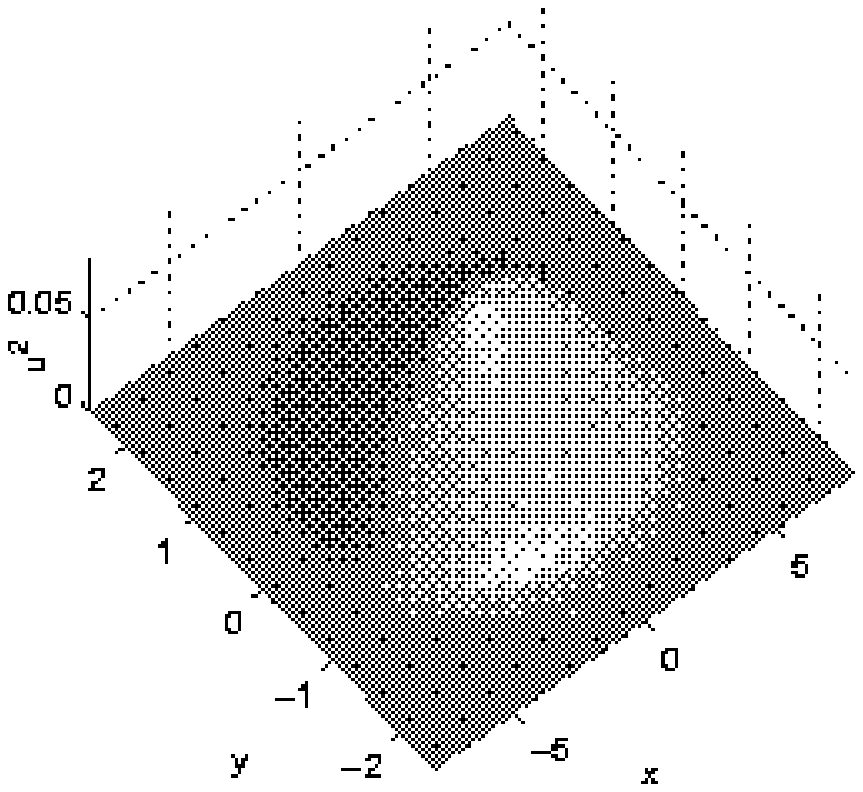,height=6.5cm,width=6cm,angle=0} 
\qquad (b).\psfig{figure=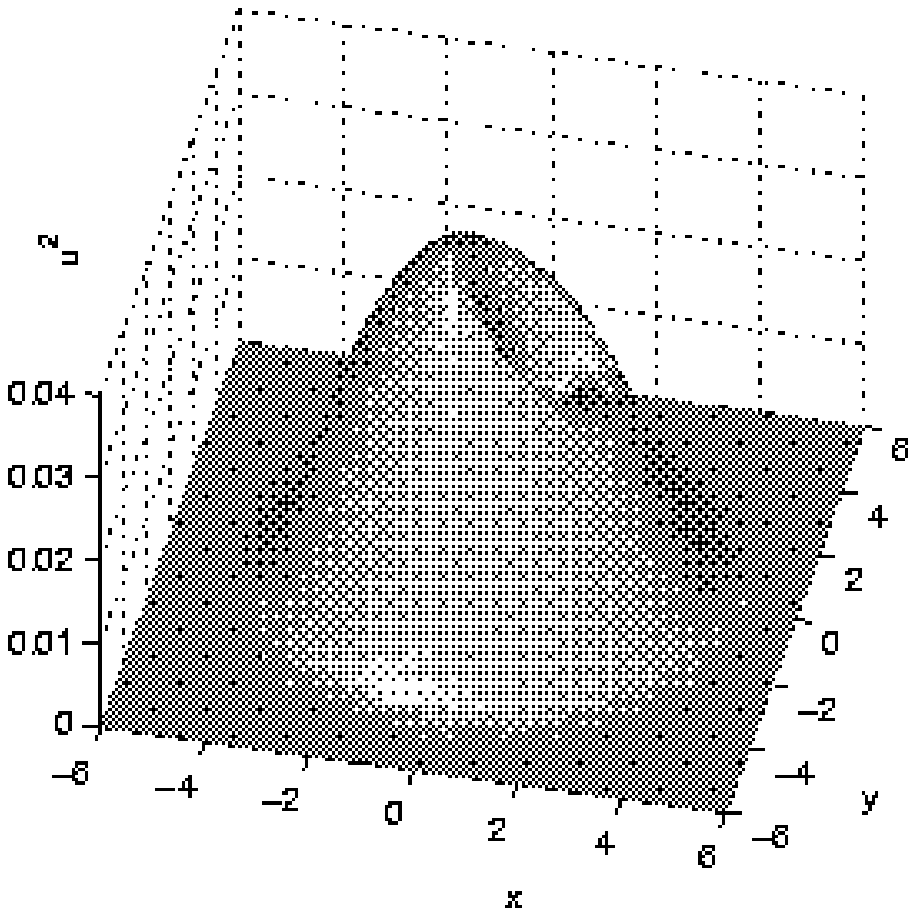,height=6.5cm,width=6cm,angle=0}} 
\caption{Surface plots of the ground state solutions $\phi_g^2(x,y)$
(labeled as $u^2$) in Example 3, case I (a),  and  case II (b).}
\end{figure}

\begin{figure}[hb] \label{fig5t}
\centerline{a).\psfig{figure=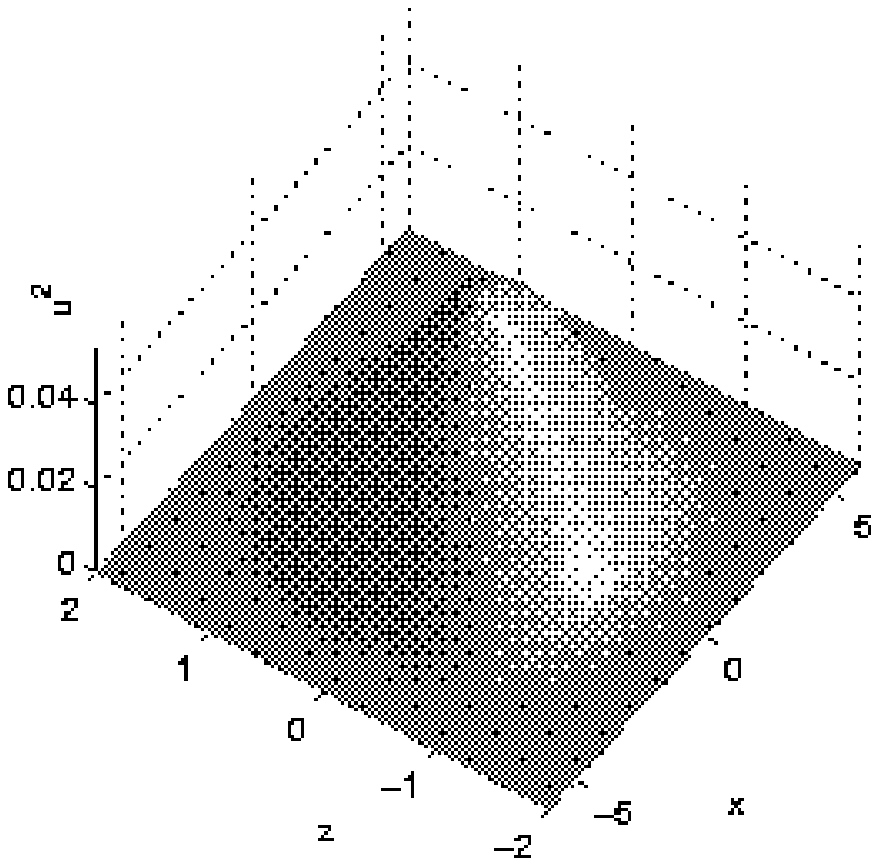,height=6.5cm,width=6cm,angle=0} 
\qquad b).\psfig{figure=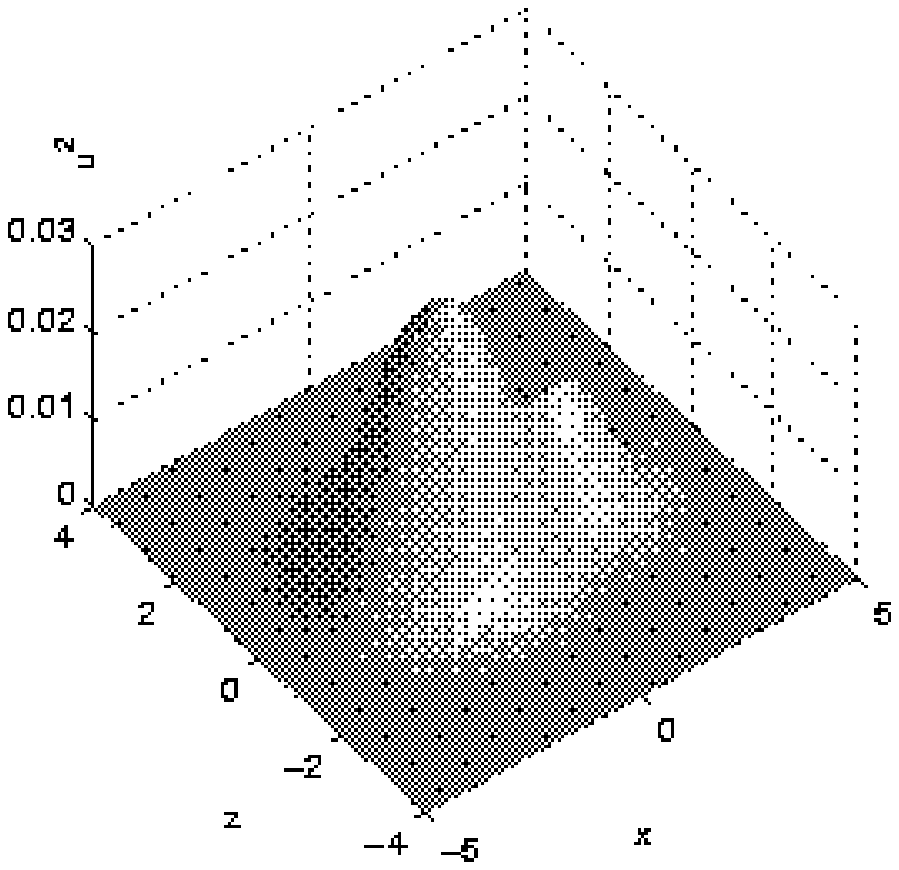,height=6.5cm,width=6cm,angle=0}} 
\caption{Surface plots of the ground state solutions $\phi_g^2(x,0,z)$
(labeled as $u^2$) in Example 4.
(a). For case I.  (b). For case II.}
\end{figure}

\begin{figure}[hb] \label{fig6t}
\centerline{(a). \psfig{figure=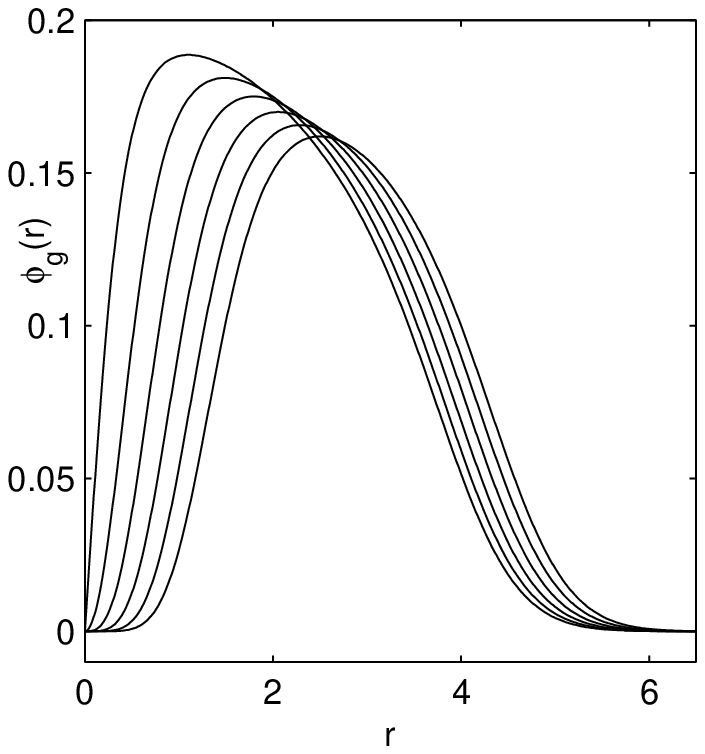,height=6.5cm,width=6cm,angle=0} 
\qquad (b). \psfig{figure=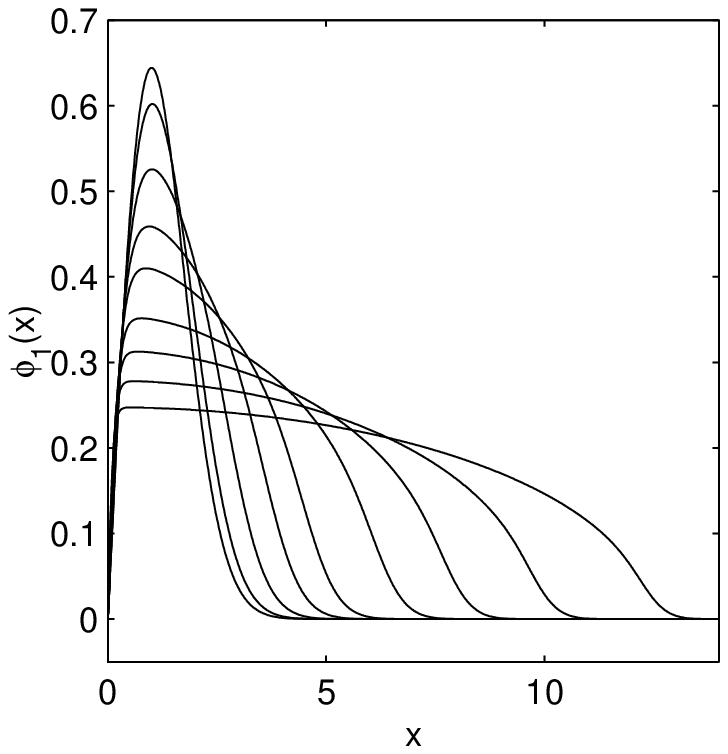,height=6.5cm,width=6cm,angle=0}}
\caption{(a). 2d central vortex states $\phi_g(r)$
in Example 5. $\bt=200$. For $m=1,\; 2,\; 3,\; 4,\; 5,\; 6$ 
(in the order of decreasing peak). 
(b). First excited state solution $\phi_1(x)$ (an odd function)
in Example 6. For $\bt=0,\; 3.1371,\; 12.5484,\; 31.371,\; 62.742,\; 
156.855,\; 313.71,\; 627.42,\; 1254.8$ (in the order of decreasing peak).}
\end{figure}

\end{document}